\documentstyle[12pt]{article}

\catcode`\@=11

\ifx\documentclass\undefined
 \def\@sect#1#2#3#4#5#6[#7]#8{\ifnum #2>\c@secnumdepth
     \let\@svsec\@empty\else
     \refstepcounter{#1}\edef\@svsec{\csname prefix#1\endcsname
        \csname the#1\endcsname\hskip 1em}\fi
     \@tempskipa #5\relax
      \ifdim \@tempskipa>\z@
        \begingroup #6\relax
          \@hangfrom{\hskip #3\relax\@svsec}{\interlinepenalty \@M #8\par}%
        \endgroup
       \csname #1mark\endcsname{#7}\addcontentsline
         {toc}{#1}{\ifnum #2>\c@secnumdepth \else
                      \protect\numberline{\csname the#1\endcsname}\fi
                    #7}\else
        \def\@svsechd{#6\hskip #3\relax  
                   \@svsec #8\csname #1mark\endcsname
                      {#7}\addcontentsline
                           {toc}{#1}{\ifnum #2>\c@secnumdepth \else
                             \protect\numberline{\csname the#1\endcsname}\fi
                       #7}}\fi
     \@xsect{#5}}
\else
    \def\@seccntformat#1{\csname prefix#1\endcsname
        \csname the#1\endcsname\quad}
\fi

\@addtoreset{equation}{section}   

\def\thebibliography#1{\section*{References\@mkboth
 {REFERENCES}{REFERENCES}}\list
 {\leftbibmark\arabic{enumi}\rightbibmark}{
 \settowidth\labelwidth{\leftbibmark #1\rightbibmark}\leftmargin\labelwidth
 \advance\leftmargin\labelsep
 \usecounter{enumi}}
 \def\newblock{\hskip .11em plus .33em minus -.07em}
 \sloppy\clubpenalty4000\widowpenalty4000
 \sfcode`\.=1000\relax}

\def\@citex[#1]#2{\if@filesw\immediate\write\@auxout{\string\citation{#2}}\fi
  \def\@citea{}\@cite{\@for\@citeb:=#2\do
    {\@citea\def\@citea{,\penalty\@m\ }\@ifundefined
       {b@\@citeb}{{\bf ?}\@warning
       {Citation `\@citeb' on page \thepage \space undefined}}%
\hbox{\csname b@\@citeb\endcsname\citemarkdelim}}}{#1}}

\def\@cite#1#2{\leftcitemark{#1 \if@tempswa , #2\fi}\rightcitemark}
\def\leftcitemark{[}
\def\rightcitemark{]}
\def\citemarkdelim{}
\def\leftbibmark{[}
\def\rightbibmark{]}

\catcode`\@=12

\def\bm#1{\mbox{\boldmath $#1$}}

\def\Frac(#1/#2){\left(\frac{#1}{#2}\right)}

\def\Beq{\begin{equation}}
\def\Eeq{\end{equation}}
\def\Beqr{\begin{eqnarray}}
\def\Eeqr{\end{eqnarray}}
\def\Beqrn{\begin{eqnarray*}}
\def\Eeqrn{\end{eqnarray*}}
\def\Bitm{\begin{itemize}}
\def\Eitm{\end{itemize}}

\begin{document}

\begin{titlepage}


\vspace{2cm}

\begin{center}\large\bf
Evolution of cosmological perturbations \\
in the universe dominated by multiple scalar fields
\end{center}

\bigskip




\begin{center}
Takashi Hamazaki\footnote{email address: yj4t-hmzk@asahi-net.or.jp}
\end{center}

\begin{center}\it
Department of Physics, Faculty of Science, Tokyo Institute of Technology,\\
Oh-okayama, Meguro, Tokyo 152-0033, Japan\\
\end{center}

\bigskip
\bigskip
\begin{center}\bf Abstract\end{center}

By efforts of several reserchers,
it is recently established that
the dynamical behavior of the cosmological perturbation on 
superhorizon scales is well approximated in terms of that 
in the long wavelength limit, 
and that the latter can be constructed 
from the evolution of corresponding exactly homogeneous universe.
Using these facts,
we investigate the evolution of the cosmological perturbation
on superhorizon scales
in the universe dominated by oscillating multiple scalar fields
which are generally interacting with each other, and 
the ratio of whose masses is incommensurable.
Since the scalar fields oscillate rapidly around the local minimum
of the potential, we use the action angle variables.
We found that this problem can be formulated as the canonical 
perturbation theory in which the perturbed part appearing as the 
result of the expansion of the universe and the interaction of 
the scalar fields is bounded by the negative power of time.
We show that by constructing the canonical transformations properly,
the transformed hamiltonian becomes simple enough to be solved.
As the result of the investigation using the long wavelength limit
and the canonical perturbation theory, under the sufficiently general 
conditions, we prove that
for the adibatic growing mode the Bardeen parameter stays constant
and that for all the other modes the Bardeen parameter decays.
From the viewpoint of the ergodic theory, it is discussed that 
as for the Bardeen parameter, the singularities appear 
probabilistically.
This analysis serves the understanding of the evolution of the 
cosmological perturbations on superhorizon scales during reheating.

PACS number(s):98.80.Cq

\end{titlepage}

\section{Introduction}

The cosmological structures 
such as the galaxies and the clusters of galaxy, etc,
are thought to be produced from the seed pertubations 
through the growth induced by the gravitational instability.
In the inflationary paradigm, the quantum fluctuation 
of the scalar field driving the inflation called the inflaton
plays the role of the seed perturbation.
In the slow rolling phase, the seed perturbations produced 
in the Hubble horizon are streched into the superhorizon scales 
by exponential cosmic expansion, and stay out of horizon 
until the perturbation comes into horizon in the Friedman 
regime.
We can calculate the amplitude of the seed pertubations 
when the potential of the inflaton is given, and 
know the evolutionary behavior of the cosmological perturbation 
in the horizon in the Friedmann universe when 
the composition of the cosmic matter is given.
Therefore in order to obtain the knowledge of the inflationary 
universe from the statistical informations of the galaxy distribution,
we need to know how the amplitudes of pertubations 
on superhorizon scales grow   
from the first horizon crossing in the slow rolling phase until 
the second horizon crossing in the Friedmann regime.

The evolution of the cosmological perturbation on the superhorizon scales 
are thought to be simple.
Because the Bardeen parameter are conserved 
in a good accuracy, when the equation of state is regular, 
and the energy of the universe is dominated by a single component.
\cite{Bardeen.J1980}
But between the inflationary regime and the Friedmann regime, reheating 
occurs.
During reheating, since the plural scalar fields oscillate rapidly
around the local minimum of the potential, and since radiation is 
generated by the energy transfer from the scalar field oscillation,
the equation of state becomes singular periodically 
\cite{Kodama.H&Hamazaki1996}, and the entropy 
modes can appear \cite{Hamazaki.T&Kodama1996}.
Therefore the approximate conservation of the Bardeen parameter 
is not self-evident, and we need to solve the evolution of 
the cosmological 
perturbations without using the conservation of the Bardeen parameter.
In general inflationary scenarios, there exisists plural scalar fields.
\cite{multiscalar}
So in this paper, we consider the universe dominated by multiple scalar 
fields.

Recently the strong method of solving the evolution of 
cosmological perturbations on superhorizon scales is developed.
The evolutionary behavior of the cosmological perturbations
on superhorizon scales is described by that in the long 
wavelength limit in a good accuracy.
\cite{Kodama.H&Hamazaki1996}
Further the long wavelength limit of the solution of 
the evolution equation of the cosmological perturbations 
is generated from the perturbation of solution of 
the evolution equation of the background homogeneous universe. 
\cite{Taruya.A&Nambu1998}, \cite{Kodama.H&Hamazaki1998}, 
\cite{Sasaki.M&Tanaka1998}.
By using the fact, we investigate the evolution of the cosmological 
perturbation on superhorizon scales in the universe dominated 
by the oscillating multiple scalar fields interacting with 
each other.

In order to use the above method, we are required to solve the 
exactly homogeneous system accurately.
During reheating, the scalar fields oscillate rapidly around the local 
minimum of the potential.
As a method of treating dynamical system with rapidly oscillating 
degrees of freedom, we use the action angle variables, the hamiltonian 
formulation and the canonical perturbation theory.
\cite{Arnold}
In the canonical perturbation theory, the hamiltonian is decomposed 
into an unperturbed part whose solution is already known and 
an perturbation part, and by constructing the change of dynamical 
variables by which pertubation part is decreased, the system is made 
simple enough to be solved.
The dynamical degrees of freedom 
are decomposed into the slowly varying amplitudes of oscillation 
called action variables which are involutive first integrals of the 
unperturbed part and the fast increasing phase of oscillation called 
angle variables.
The angle variables are eliminated from the hamiltonian 
by the canonical transformation.
By using the method of the long wavelength limit and the canonical
perturbation theory, we investigate the evolutionay behavior of the
Bardeen parameter.
While we adopt the action angle variables as the dynamical variables
in the viewpoint of treating the reheating, from the standpoint
decomposing the adiabatic and the entropy perturbation, the different
variables were proposed.
\cite{adivsentropy}
Also in other approach the multiple scalar fields system was
investigated.
\cite{otherapproach}

This paper is organized as follows.
In chapter 2, using the result of 
the paper \cite{Kodama.H&Hamazaki1998}, 
we explain the way to 
construct a solution of the perturbation equation 
in the long wavelength limit
from a perturbation of a solution of the exactly homogeneous 
universe.
In chapter 3, we write the system of the multiple scalar fields 
in the homogeneous universe in the hamiltonian form and fix 
the model which we treat concretely.
In chapter 4, we apply the canonical perturbation theory
to our system.
The scalar fields oscillation is decomposed into the action 
angle variables, the former of which and the latter of which 
represent the slowly varying amplitudes 
and fast increasing phases of the 
scalar fields oscillation, respectively.
By using the canonical transformations reccurently, we decrease 
the power of time variable which the interaction parts contain 
and eliminate fast oscillation from the evolution equation and 
extract the slow secular motion.
We give the evaluation of the generating functions of the 
canonical transformations by which the transformed system becomes
simple enough to be solved.  
In chapter 5, we investigate the evolution of the multiple scalar 
fields in the homogeneous universe.
We prove the fact that the action variables are perpetually stable.
In chapter 6, we investigate the evolution 
of the cosmological perturbations in the long wavelength limit
paying attention to that of the Bardeen parameter.
The upper bound of evolution of the perturbation variables and 
the uniqueness of the growing mode are derived.
As a result, it is concluded that the Bardeen parameter stays
constant in a good accuracy when the masses of the scalar fields
are incommensurable.
In this case, the Bardeen parameter becomes singular 
probabilistically.
So in terms of the ergodic theory, we discuss the distribution
of the singularity of the Bardeen parameter.
The chapter 7 is devoted to the summary and the conclusion.

Throughout the paper the natural units $c=\hbar=1$ are adopted and
$8\pi G$ is denoted as $\kappa^2$. Further the notations for the
perturbation variables adopted in the article
\cite{Kodama.H&Sasaki1984} are used and their definitions are
sometimes omitted except for those newly defined in this paper

\section{Review; Evolution of cosmological perturbation in the 
longwavelength  limit}

The dynamical behavior of cosmological perturbations on superhorizon 
scales 
is well approximated by that of cosmological perturbations in the long 
wavelength limit \cite{Kodama.H&Hamazaki1996},
and the evolution of cosmological perturbations 
in the $k \rightarrow 0$ limit
can be constructed from the perturbation 
of the exactly homogeneous solution
\cite{Kodama.H&Hamazaki1998}.
In this section, we present the results of the paper 
\cite{Kodama.H&Hamazaki1998} in case of the universe dominated 
by multi-component scalar fields.
Next we explain two way to construct the perturbation 
of the exactly homogeneous solution.

We only consider perturbations on a spatially flat($K=0$) Robertson-Walker 
universe throughout the paper. Hence the background metric is given by
\begin{equation}
ds^2 = -dt^2 + a(t)^2 d\bm{x}^2.
\end{equation}
We consider the universe dominated by multi-component scalar fields, 
and the energy-momentum tensor is given by 
\begin{equation}
T^\mu_\nu=\nabla^\mu\phi\cdot\nabla_\nu\phi
-{1\over2}\delta^\mu_\nu\left(\nabla^\lambda\phi\cdot\nabla_\lambda\phi + 2U
\right).
\end{equation}
In the long wavelength limit, the evolution of the gauge invariant
variable $Y_i$ representing the fluctuation of the scalar field 
$\phi_i$ in the flat time slice are described as
\begin{eqnarray}
 Y_i &=& {\mit \chi_i} + {\dot\phi_i \over H}\int dt{H^2\over 2U}
         W\left({\dot\phi\over H},{\mit \chi}\right),
\nonumber \\
 W(X_1,X_2)&:=& X_1\cdot \dot X_2 - \dot X_1\cdot X_2,
\label{fieldperturbationformula}
\end{eqnarray}
where ${\mit \chi}_i$ is a combination of exactly homogeneous perturbation 
variables described as
\begin{equation}
 {\mit \chi}_i = \delta \phi_i - {\dot\phi_i \over H} {\delta a \over a}.
\end{equation}
In the $k=0$ limit, it is noticed that the equality
\begin{equation}
{a^3H^2\over U}W\left({\dot\phi\over H},{\mit \chi}\right)={\mathrm const}
\label{k0limitWronskian}
\end{equation}
holds.

Corresponding to the way how the homogeneous solution is written,
we can construct the exactly homogeneous perturbation in the different 
two ways.
One is the way how it is given as derivative of the homogeneous solution
with respect to the solution constants.
The homogeneous solution is described as 
\begin{equation}
 \phi_i = \phi_i (t, C) ,~~~~a=a(t, C),
\end{equation}
or 
\begin{equation}
 \phi_i = \phi_i (a, C),
\end{equation}
where $C$ is the set of all the solution constants
characterizing homogeneous solution completely.
${\mit \chi}_i$ is expressed as
\begin{equation}
 {\mit \chi}_i = {\partial \phi_i (t, C) \over \partial C}
 - {\dot\phi_i \over H} {1 \over a} {\partial a (t, C) \over \partial C}
\end{equation}
in the above case, and is expressed as
\begin{equation}
 {\mit \chi}_i = {\partial \phi_i (a, C) \over \partial C}
\end{equation}
in the below case.
Noticing the functional relation $a=a(t, C)$,
it is seen that the above two expressions are exactly the same.
In this paper, the second expression is used when the evolution
of the cosmological perturbations in the long wavelength limit
is investigated.

On the other hand, the solution is often expressed as 
union of the manifold on which the first integral is constant
\begin{equation}
F_n(\phi,p_{\phi},a) = {\rm constant}~~~
 (n = 1, \cdot \cdot \cdot, 2N)
\end{equation}
where $p_{\phi}$ is canonical momentum conjugate to $\phi$
as defined later.
In this case, the exactly homogeneous perturbation is produced 
as
\begin{equation}
 {\mit \chi}_i = \{F_n(\phi,p_{\phi},a), \phi_i\} =
- {\partial F_n(\phi,p_{\phi},a) \over \partial p_{\phi_i}}.
\end{equation}

\section{Evolution equations of corresponding exactly homogeneous 
universe}

In this section, we formulate the evolution of the exactly 
homogeneous universe dominated by multiple scalar fields 
as a hamiltonian form.
Since our system are derived from the general covariant theory,
it becomes a constrained system with gauge degree of freedom.
By choosing the function of the scale factor as the time variable,
we fix the gauge.
It is the important first step since the evolution of the 
cosmological perturbations in the long wavelength limit is generated
from the exactly homogeneous perturbations.

We consider the situation where the multiple scalar fields interact 
with each other
\begin{equation}
 S = \int dt \bigl( p_a \dot{a} + \sum_i p_{\phi_i} \dot{\phi_i} 
     - N C_H \bigr). 
\end{equation}
Here $N$ is the lapse function and $C_H$ is the Hamiltonian constraint
written as
\begin{equation}
 C_H = \Omega a^3 \Bigl( - {\kappa^2 \over 12} {1 \over \Omega^2 a^4}
       p^2_a + {1 \over 2} {1 \over \Omega^2 a^6} \sum_i p_{\phi_i}^2
       + U(\phi) \Bigr),
\end{equation}
where $\Omega$ is the comoving volume of the constant time hypersurface.
The potential $U(\phi)$ is decomposed as
\begin{equation}
 U(\phi)= {1 \over 2} \sum_i m_i^2 \phi_i^2 + U_{\rm int}(\phi).
\end{equation}
where $U_{\rm int}$ is assumed to be a sum of monomials of $\phi$ of
degree not smaller than $3$ as in most physical context.
We fix the gauge by 
\begin{equation}
 0 = \psi = a - t,
\end{equation}
and its consistency condition with time evolution determines $N$ as
\begin{equation} 
 0 = {d \psi \over dt} = - {\kappa^2 \over 6} {N \over \Omega a}
 p_a - 1.
\end{equation}
For notational simplicity we write $a$ in stead of $t$ henceforth.
By using Dirac bracket formalism, the Hamiltonian constraint $0 =  C_H$
is changed into a strong equation, and $p_a$ can be removed away.
We obtain a reduced action with no constraint 
\begin{eqnarray}
 S &=& \int \sum_i p_{\phi_i} d \phi_i - h_a da,\\
 h_a &=& { 2 \sqrt{3} \over \kappa } \Omega a^2 \Bigl(
         {1 \over 2} {1 \over \Omega^2 a^6} \sum_i p_{\phi_i}^2
         + U(\phi) \Bigr)^{1/2}.
\end{eqnarray}
We make all physical quantities non-dimensional by dividing them 
with typical scales as 
\begin{eqnarray}
 \zeta &=& \bigl( {a \over a_0} \bigr)^{3/2},\\
 \Phi &=& {\phi \over \phi_0},\\
 P_{\phi} &=& {p_{\phi} \over a_0^3 m_0 \phi_0 \Omega},\\
 \mu_i &=& {m_i \over m_0},\\
 \epsilon &=& {\sqrt{3} \over 2} \kappa \phi_0,\\
 S^{'} &=& {S \over \Omega a_0^3 m_0 \phi_0^2},\nonumber \\ 
       &=&  \int \sum_i P_{\phi_i} d \Phi_i 
\nonumber \\
       && 
            - {2 \over \epsilon} 
            \Bigl(
    {1 \over 2} {1 \over \zeta^2} \sum_i P_{\phi_i}^2 
    + {1 \over 2} \zeta^2 \sum_i \mu_i^2 \Phi_i^2 
    + \zeta^2 {1 \over m_0^2 \phi_0^2 }U_{\rm int}( \phi_0  \Phi)
            \Bigr)^{1/2} d\zeta.
\end{eqnarray}

From now on, we use $t$ instead of $\zeta$ 
without possibility of confusion.
In this paper, the evolution of the cosmological perturbations
in the long wavelength limit is generated from the derivative
of the exactly homogeneous solution with respect to solution 
constant with the scale factor fixed.
Since we adopt the function of the scale factor as the time variable,
in order to generate the evolution equations followed by the exactly
homogeneous perturbations, it is sufficient to take variations with
respect to the canonical variables with the time variable fixed
in the unperturbed canonical equations of motion.

\section{ Integration Method }

This section is the main part of the paper.
Since the scalar fields oscillate around the local minimum of the 
potential, we use the action angle variables, the former of which 
and the latter of which represent the slowly varying amplitudes 
and the fast increasing phases of scalar fields oscillation, 
respectively.
We found that our system can be formulated as the canonical 
perturbation theory the perturbed part of which depending on 
the angle variables and the negative power of the time appears 
as a result of the expansion of the universe and the interaction 
of the scalar fields.
Reducing the evolution equation into 
a simpler form whose perturbed part is sufficiently small
by constructing appropriate coordinate 
transformations is known as the method of variation
of constants mathematically.
Since our system is described in the hamiltonian form,
we use the canonical transformations.
By constructing the canonical transformation properly,
we show that the perturbed parts of the hamiltonian can be decreased
so that the transformed hamiltonian can be solved easily. 

The starting hamiltonian is given by
\begin{equation}
    H =  {2 \over \epsilon} 
            \Bigl(
    {1 \over 2} {1 \over t^2} \sum_i P_{\phi_i}^2 
    + {1 \over 2} t^2 \sum_i \mu_i^2 \Phi_i^2 
    + t^2 {1 \over m_0^2 \phi_0^2 }U_{\rm int}( \phi_0  \Phi)
            \Bigr)^{1/2}.
\end{equation}
By performing canonical transformation 
$(\Phi_i, P_{\phi_i}) \rightarrow (\xi^i, \eta^i)$
induced by generating function as
\begin{equation}
 W(\Phi, \eta) = \sum_i \sqrt{2} \sqrt{\mu_i} t \eta^i \Phi_i +
   \sum_i i \mu_i t^2 {1 \over 2} \Phi_i^2 - 
   \sum_i i {1 \over 2} \eta^{i2},
\end{equation}
we introduce complex canonical variables as
\begin{eqnarray}
 &&
 \xi^i = {1 \over \sqrt {2}} 
   (\sqrt{\mu_i} t \Phi_i - {1 \over \sqrt{\mu_i}} {1 \over t} i P_{\phi_i}),
\\
 &&
 \eta^i = {1 \over \sqrt {2}} 
  (- i \sqrt{\mu_i} t \Phi_i + {1 \over \sqrt{\mu_i}} {1 \over t} P_{\phi_i}).
\end{eqnarray}
Then the transfomed hamiltonian is given by
\begin{eqnarray}
 H' &=& H + {\partial W \over \partial t}
\nonumber \\
 &=&
 {2 \over \epsilon} 
            \Bigl(
    i \sum_i \mu_i \xi^i \eta^i
    + t^2 {1 \over m_0^2 \phi_0^2 }U_{\rm int}
            \Bigr)^{1/2}
    + {i \over 2} {1 \over t} 
            \Bigl(
    \sum_i \eta^{i2} + \sum_i \xi^{i2} 
            \Bigr)
.
\end{eqnarray}
Since the interaction term $U_{\rm int}$ is written as a sum of 
monomials of $\Phi_i$;
\begin{equation}
 \Phi_i = {1 \over \sqrt {2}} 
   ( {1 \over \sqrt{\mu_i}} {1 \over t} \xi^i + 
   i {1 \over \sqrt{\mu_i}} {1 \over t} \eta^i ),
\end{equation}
of degree not smaller than $3$,
we assume that $U_{\rm int}$ is a sum of monomials in terms of 
$\zeta^i / t=(\xi^i / t, \eta^i / t)$ of power not smaller than $3$ 
with $t$-independent coefficients;
\begin{equation}
 {t^2 \over m_0^2 \phi_0^2} U_{\rm int} =
    \lambda t^2 \sum_{|k|+|l| = 3} c_{kl} ({\xi \over t})^k
   ({\eta \over t})^l
    + \lambda t^2 \sum_{|k|+|l| \ge 4} c_{kl} ({\xi \over t})^k
   ({\eta \over t})^l,
\end{equation}
where $\lambda$ is a constant characterizing the strength of interaction.
We introduce the action angle variables by
\begin{eqnarray}
 &&
 \xi^i = \sqrt{J^i} e^{i \theta^i},
\\
 &&
 \eta^i = - i \sqrt{J^i} e^{- i \theta^i}.
\end{eqnarray}
Since $k \ne l$ under the condition $|k|+|l|=3$, the first term of the 
right hand side $K_3$;
\begin{equation}
 K_3 = {\lambda \over t} k_3 (J, \theta),
\end{equation}
where $k_3$ is $2 \pi$-periodic in 
$\theta = (\theta^1, \cdot \cdot \cdot, \theta^N)$ and
is of power $3 / 2$ in terms of $J$, satisfies 
\begin{equation}
 {1 \over (2 \pi)^N} \int^{2 \pi}_0 \cdot \cdot \cdot 
 \int^{2 \pi}_0 d^N \theta K_3 (J, \theta, t) = 0.
\end{equation}
The second term $K_4$ is written as
\begin{equation}
 K_4 = \lambda \sum^{\infty}_{p=2} {1 \over t^p} 
 k_{4~p+2} (J, \theta),
\end{equation}
where $k_{4~p+2}$ is $2 \pi$-periodic in 
$\theta = (\theta^1, \cdot \cdot \cdot, \theta^N)$ and
is of power $(p+2) / 2$ in terms of $J$. 
The transformed hamiltonian $H'$ is rewritten as
\begin{eqnarray}
 H &=& {2 \over \epsilon} (\mu \cdot J + K_3 + K_4)^{1 / 2}
    - \sum_i {J^i \over t} \sin{2 \theta^i}
\nonumber \\
   &=& {2 \over \epsilon} (\mu \cdot J)^{1 / 2} + {\tilde H}.
\end{eqnarray}
The interaction part ${\tilde H}$ is decomposed as
\begin{eqnarray}
 {\tilde H} &=& {2 \over \epsilon} (\mu \cdot J + K_3 + K_4)^{1 / 2}
                - {2 \over \epsilon} (\mu \cdot J)^{1 / 2}
                - \sum_i {J^i \over t} \sin{2 \theta^i}
\nonumber \\
 &=& A'_1 + B'_1,
\end{eqnarray}
where 
\begin{eqnarray}
 && 
 A'_1 = {1 \over \epsilon} {1 \over (\mu \cdot J)^{1 / 2}} K_4
   + {2 \over \epsilon} (\mu \cdot J)^{1 / 2} \sum^{\infty}_{n=2}
  \pmatrix{
  {1 / 2} \cr
  n \cr }
  ({K_3 + K_4 \over \mu \cdot J})^n,
\\
 &&
 B'_1 = {1 \over \epsilon} {1 \over (\mu \cdot J)^{1 / 2}} K_3
   - \sum_i {J^i \over t} \sin{2 \theta^i}.
\end{eqnarray}
Now we assume that $\lambda / \epsilon$ is at most of order unity.
By transferring $\theta$-dependent part of $A'_1$ to $B'_1$,
we obtain as a starting hamiltonian
\begin{equation}
  H = {2 \over \epsilon} (\mu \cdot J_1)^{1 / 2} 
     + A_1 (J_1, t) + B_1 (J_1, \theta_1, t),
\label {startinghamiltonian}
\end{equation}
satisfying  
(i) $\mu \cdot J_1$ is bounded as
\begin{equation}
 |\mu \cdot J_1| \ge \sigma_1,
\end{equation}
for some positive $\sigma_1$ on the domain 
$(J_1,~1/t) \in D'_1(\rho_1)$ where 
\begin{equation}
 D'_n(\rho) = \{ (J, {1 \over t}); 
             |J-J_{n0}| \le \rho,  
 |{1 \over t}| \le 1+\rho, |{\rm Im}{1 \over t}| \le \rho \}.
\end{equation}
where $J_0$ is the initial value of $J$; $J_0 \equiv J(t=1)$.

(ii) $t^2 A_1$ is analytic in terms of $J_1,~1/t$ and is real for real 
$J_1,~1/t$ and the inequality 
\begin{equation}
 |t^2 A_1| \le M^{(1)}_1 
\end{equation}
holds for some positive $M^{(1)}_1$ on the domain $D_1(\rho_1)$, where

\begin{equation}
 D_n(\rho) = \{ (J, \theta, {1 \over t}); 
             |J-J_{n0}| \le \rho, 
 |{\rm Im} \theta| \le \rho,
 |{1 \over t}| \le 1+\rho, |{\rm Im}{1 \over t}| \le \rho \}.
\end{equation}

(iii) $t B_1$ is analytic in terms of $J_1,~\theta_1,~1/t$ 
and is real for real $J_1,~\theta_1,~1/t$ and is $2 \pi$-periodic in 
$\theta_1  = (\theta^1_1, \cdot \cdot \cdot, \theta^N_1)$ 
and the inequality 
\begin{equation}
 |t B_1| \le M^{(1)}_2 
\end{equation}
holds for some positive $M^{(1)}_2$ and satisfy
\begin{equation}
 {1 \over (2 \pi)^N} \int_0^{2 \pi} \cdot \cdot \cdot \int_0^{2 \pi}
 d^N \theta_1 t B_1 (J_1, \theta_1, t)=0,
\end{equation}
on the domain $D_1(\rho_1)$.

It is noticed that the perturbed parts $A_1$, $B_1$ appearing as the 
result of the expansion of the universe and the interactions of the 
scalar fields depend on the time as negative power.
Since we are interested in the evolution for sufficiently large $t$,
this property plays an essential role in the following analysis.

In reheating phase, scalar fields oscillate rapidly.
Since rapid oscillation is diffcult to treat,
we construct change of coordinates in order to eliminate
fast phases called the angle variables 
from the transformed equations of perturbed motion.
The canonical perturbation theory is effective for this purpose, 
and can separate the slow secular motions from the fast oscillations
by performing the canonical transformations recurrently,
when the frequencies of unperturbed motion are incommensurable.
With noticing that we can eliminate the dependence 
on the angle variables from the hamiltonian, it is important to 
understand that we can increase the power of the inverse of the 
time $1 / t$ which is contained by the perturbed part 
depending on the angle variables $B$ since we are interested in 
the evolution for sufficiently large $t$.

{\bf Proposition1 (Theorem for the canonical transformations)}

Suppose that $m \ge 1$.
Suppose that the hamiltonian 
\begin{equation}
 H^{(m)} = {2 \over \epsilon} (\mu \cdot J_m)^{1 / 2} 
             + A_m (J_m, t) + B_m (J_m, \theta_m, t)
\end{equation}
satisfy the following conditions:

(i) For some positive $C$ and $d$, the inequality
\begin{equation}
 |k \cdot \mu| \ge C {1 \over |k|^{d} }
\end{equation}
holds for every integral-valued vector $k=(k_1, \cdot \cdot \cdot, k_N)$.

(ii) $\mu \cdot J_m$ is bounded as
\begin{equation}
 |\mu \cdot J_m| \ge \sigma_m,
\end{equation}
for some positive $\sigma_m$
on the domain $(J_m, {1 / t}) \in D'_m(\rho_m)$.

(iii) $t^2 A_m(J_m, t)$ is analytic and real for real $J_m$, $1 / t$ 
and the inequality 
\begin{equation}
 |t^2 A_m| \le M^{(m)}_1
\end{equation}
holds for some positive $M^{(m)}_1$ on the domain 
$(J_m, {1 / t}) \in D'_m(\rho_m)$.

(iv) $t^m B_m(J_m, \theta_m, t)$ is periodic in 
$\theta_m = (\theta^{(m)}_1, \cdot \cdot \cdot, \theta^{(m)}_N)$.
It is analytic and real for real $J_m$, $\theta_m$, $1 / t$ 
and the inequality 
\begin{equation}
 |t^m B_m| \le \epsilon^{m - 1} M^{(m)}_2
\end{equation}
holds for some positive $M^{(m)}_2$ 
on the domain $(J_m, \theta_m, {1 / t}) \in D_m(\rho_m)$
and satisfy
\begin{equation}
 {1 \over (2 \pi)^N} \int_0^{2 \pi} \cdot \cdot \cdot \int_0^{2 \pi}
 d^N \theta_m t^m B_m (J_m, \theta_m, t)=0.
\end{equation}

For an arbitrary positive $\delta$,
there exists $\epsilon_0$ such that, for an arbitrary $\epsilon$
satisfying
\begin{equation}
 0 < \epsilon < \epsilon_0,
\end{equation}
there exists the canonical transformation induced by the generating function
$S_m (J_{m+1}, \theta_m, t)$ satisfying the following condition;
 
(v) $t^m S_m(J_{m+1}, \theta_m, t)$ is periodic in 
$\theta_m = (\theta^{(m)}_1, \cdot \cdot \cdot, \theta^{(m)}_N)$.
It is analytic and real for real $J_{m+1}$, $\theta_m$, $1 / t$ 
and the inequality 
\begin{equation}
 |t^m S_m| \le \epsilon^m L^{(m)}_1
\end{equation}
holds for some positive $L^{(m)}_1$ 
on the domain $(J_{m+1}, \theta_m, {1 / t}) \in D_{m+1}(\rho_{m+1})$, 
where $\rho_{m+1} = \rho_m - \delta$.

The transformed system defined by 
\begin{eqnarray}
&&
 J_m = J_{m + 1} + {\partial S_m \over \partial \theta_m},
\\
&&
 \theta_{m + 1} = \theta_m + {\partial S_m \over \partial J_{m+1}},
\\
&&
 H^{(m+1)} = H^{(m)} + {\partial S_m \over \partial t}
\\
&& ~~~~~~ = {2 \over \epsilon} (\mu \cdot J_{m+1})^{1 / 2} 
             + A_{m+1} (J_{m+1}, t) 
             + B_{m+1} (J_{m+1}, \theta_{m+1}, t)
\end{eqnarray}
has the following properties;

(vi)$\mu \cdot J_{m+1}$ is bounded as
\begin{equation}
 |\mu \cdot J_{m+1}| \ge \sigma_{m+1},
\end{equation}
for some positive $\sigma_{m+1}$
on the domain $(J_{m+1}, {1 / t}) \in D'_{m+1}(\rho_{m+1})$

(vii) $t^2 A_{m+1}(J_{m+1}, t)$ is analytic and real for real 
$J_{m+1}$, $1 / t$ and the inequality 
\begin{equation}
 |t^2 A_{m+1}| \le M^{(m+1)}_1
\end{equation}
holds for some positive $M^{(m+1)}_1$ on the domain 
$(J_{m+1}, {1 / t}) \in D'_{m+1}(\rho_{m+1})$.

(viii) $t^{m+1} B_{m+1}(J_{m+1}, \theta_{m+1}, t)$ is periodic in 
$\theta_{m+1} = (\theta^{(m+1)}_1, \cdot \cdot \cdot, \theta^{(m+1)}_N)$.
It is analytic and real for real $J_{m+1}$, $\theta_{m+1}$, $1 / t$ 
and the inequality 
\begin{equation}
 |t^{m+1} B_{m+1}| \le \epsilon^m M^{(m+1)}_2
\end{equation}
holds for some positive $M^{(m)}_2$ 
on the domain $(J_{m+1}, \theta_{m+1}, {1 / t}) \in D_{m+1}(\rho_{m+1})$ 
and satisfy 
\begin{equation}
 {1 \over (2 \pi)^N} \int_0^{2 \pi} \cdot \cdot \cdot \int_0^{2 \pi}
 d^N \theta_{m+1} t^{m+1} B_{m+1} (J_{m+1}, \theta_{m+1}, t)=0.
\end{equation}

{\it Proof}

We consider the canonical transformation induced by the generating 
function given by
\begin{equation}
 S_m (J_{m+1}, \theta_m, t) =
 \sum_{k \ne 0} S_k (J_{m+1}, t) e^{i k \cdot \theta_m}.
\end{equation}
$B_m$ is decomposed as 
\begin{equation}
 B_m (J_{m+1}, \theta_m, t) = 
 \sum_{k \ne 0} b_k (J_{m+1}, t) e^{i k \cdot \theta_m}. 
\end{equation}
The transformed hamiltonian is 
\begin{eqnarray}
 &&
 H^{(m+1)} = H^{(m)} + {\partial S_m \over \partial t}
\nonumber \\
 &&
\quad \quad \quad
  = {2 \over \epsilon} (\mu \cdot J_{m+1})^{1 / 2}
   + A_m (J_{m+1}, t) 
\nonumber \\
 &&
 \quad \quad \quad \quad
  + {1 \over  \epsilon} {1 \over (\mu \cdot J_{m+1})^{1 / 2}} 
    \mu \cdot {\partial S_m (J_{m+1}, \theta_m, t) 
               \over \partial \theta_m }
  + B_m (J_{m+1}, \theta_m, t)
\nonumber \\
 &&
\quad \quad \quad \quad
 + R_1 + R_2 + R_3 +
 {\partial S_m (J_{m+1}, \theta_m, t)
  \over \partial t }
 \end{eqnarray}
where $R_1$, $\cdot \cdot \cdot$ $R_3$ are defined as 
\begin{eqnarray}
 &&
 R_1 = {2 \over \epsilon} (\mu \cdot J_m)^{1 / 2}
       - {2 \over \epsilon} (\mu \cdot J_{m+1})^{1 / 2}
       - {1 \over \epsilon} 
         {1 \over (\mu \cdot J_{m+1})^{1 / 2}} 
    \mu \cdot {\partial S_m (J_{m+1}, \theta_m, t) 
               \over \partial \theta_m },
\\
 &&
 R_2 = A_m (J_m, t) - A_m (J_{m+1}, t),
\\
 &&
 R_3 = B_m (J_m, \theta_m, t) - B_m (J_{m+1}, \theta_m, t).
\end{eqnarray}
We determine the generating function $S_m$ so that 
the leading term of the oscillating part can be eliminated;
\begin{equation}
    {1 \over \epsilon} {1 \over (\mu \cdot J_{m+1})^{1 / 2}} 
    \mu \cdot {\partial S_m (J_{m+1}, \theta_m, t) 
               \over \partial \theta_m } =
    - B_m (J_{m+1}, \theta_m, t).
\end{equation}
By comparing the fourier components in the both side,  
we obtain
\begin{equation}
 S_k (J_{m+1}, t) = i \epsilon (\mu \cdot J_{m+1})^{1 / 2}
 {1 \over (\mu \cdot k)} b_k (J_{m+1}, t).
\end{equation}
Since from (iv), on the domain 
$(J_{m+1}, {1 / t}) \in D'_m (\rho_m)$, the inequality 
\begin{equation}
 |t^m b_k (J_{m+1}, t)| \le \epsilon^{m-1} M^{(m)}_2 
                        \exp( - |k| \rho_{m-1} )
\end{equation}
holds,
we evaluate the upper bound of the fourier components $S_k$;
\begin{eqnarray}
 |t^m S_k (J_{m+1}, t)| &\le& \epsilon 
 (\mu \cdot J_{m0} + N \mu \rho_m)^{1 / 2}
 {|k|^d \over C} \epsilon^{m-1} M^{(m)}_2 
                        \exp( - |k| \rho_m )
\nonumber \\
 &\le& 
 \epsilon^m 
 (\mu \cdot J_{m0} + N \mu \rho_m)^{1 / 2} 
 {M^{(m)}_2 \over C} ({d \over e})^d {1 \over \delta^d}
 \exp \{ - |k| (\rho_m -\delta) \}.
\end{eqnarray}
where in proceeding from the first row to the second row
we use the inequality 
\begin{equation}
 |k|^d \le ({d \over e})^d 
 {\exp(|k| \delta) \over \delta^d},
\end{equation}
for arbitrary positive $|k|$, $d$, $\delta$.
On the domain $|J_{m+1} - J_{m0}| \le \rho_m$,
$|{\rm Im} \theta_m| \le \rho_m - 2 \delta$,
$t^m S_m$ is bounded as 
\begin{eqnarray}
 |t^m S_m| &\le& \sum_{k \ne 0} | t^m S_k | 
 \exp(|k| |{\rm Im}\theta_m|)
\nonumber \\
 &\le&
 \epsilon^m (\mu \cdot J_{m0}+N \mu \rho_m)^{1 / 2} 
 { M^{(m)}_2 \over C }
 ({d \over e})^d {1 \over \delta^d} 
 \sum_{k \ne 0} \exp(- |k| \delta)
\nonumber \\
 &\le&
 \epsilon^m L^{(m)}_1,
\label {upperboundofgeneratingfn}
\end{eqnarray}
where $L^{(m)}_1$ is defined as
\begin{equation}
 L^{(m)}_1 = (\mu \cdot J_{m0}+N \mu \rho_m)^{1 / 2} 
 { M^{(m)}_2 \over C }
 ({d \over e})^d {1 \over \delta^d} ({4 \over \delta})^N
\end{equation}
In passing from the second row to the third row,
we use the inequality 
\begin{eqnarray}
 \sum_{k \ne 0} \exp(- |k| \delta) &<&
     \bigl[
 1 + 2 \sum_{k > 0} \exp(-k \delta)
     \bigr]^N
\nonumber \\
 &<& ({4 \over \delta})^N,
\end{eqnarray}
where we assume that $\delta < 1$.
On the domain $|J_{m+1} - J_{m0}| \le \rho_m$,
$|{\rm Im} \theta_m| \le \rho_m - 3 \delta$,
$|{1 / t}| \le 1 + \rho_m$,
$t^m {\partial S_m / \partial \theta_m}$ is bounded as
\begin{equation}
 | t^m 
 {\partial S_m (J_{m+1}, \theta_m, t) \over \partial \theta_m}|
 \le \epsilon^m {L^{(m)}_1 \over \delta}.
\end{equation}
So the inequality
\begin{equation}
 |J_{m0} - J_{m+1~0}|=
|{\partial S_m (J_{m+1~0}, \theta_{m~0}, t=1) \over \partial \theta_{m~0}}|
\le \epsilon^m {L^{(m)}_1 \over \delta}
\end{equation}
holds.
Of course, the above two inequalities hold on the domain 
$|J_{m+1} - J_{m0}| \le \rho_m - 3 \delta$,
$|{\rm Im} \theta_m| \le \rho_m - 3 \delta$,
$|{1 / t}| \le 1 + \rho_m$.
Then 
\begin{eqnarray}
 |J_{m+1} - J_{m+1~0}| &\le& |J_{m+1} - J_{m0}| + |J_{m0} - J_{m+1~0}|
\nonumber \\
 &\le& \rho_m - 3 \delta + \epsilon^m {L^{(m)}_1 \over \delta}.
\end{eqnarray}
When we assume that $\epsilon$ satisfies 
\begin{equation}
 \epsilon^m {L^{(m)}_1 \over \delta} \le \delta,
\end{equation}
the inequality
\begin{equation}
 |J_{m+1} - J_{m+1~0}| \le \rho_m - 2 \delta
\end{equation}
holds. The inequality (\ref {upperboundofgeneratingfn}) holds on the 
domain $(J_{m+1}, \theta_m, {1 / t}) \in D(\rho_m - 2 \delta)$.

Now for an arbitrarily fixed $J_{m+1}$ on the domain 
$|J_{m+1} - J_{m+1~0}| \le \rho_m - \delta$,
we consider the mapping $T_{J_m+1}$ defined by
\begin{equation}
 T_{J_{m+1}}; \theta_m \rightarrow \theta_{m+1} = 
 \theta_m 
 + {\partial S_m \over \partial J_{m+1}}(J_{m+1}, \theta_m, t)
\end{equation}
For an arbitrary $\theta_m$ on the domain 
$|{\rm Im} \theta_m| \le \rho_m - 2 \delta$,
the inequality 
\begin{eqnarray}
 |\theta_m - \theta_{m+1}| &\le& 
 |{\partial S_m \over \partial J_{m+1}}(J_{m+1}, \theta_m, t)|
\nonumber \\
 &\le&
 \epsilon^m (1 + \rho_m)^m {L^{(m)}_1 \over \delta}
\end{eqnarray}
holds.
Therefore the mapping $T_{J_{m+1}}$ is injective (so diffeomorphic)
on the domain 
$|{\rm Im} \theta_m| \le \rho_m-2\delta
-4 \epsilon^m (1 + \rho_m)^m L^{(m)}_1 / \delta$.
$T_{J_{m+1}}$ maps an arbitrary $\theta_m$ on the domain 
$|{\rm Im} \theta_m| \le \rho_m - 2 \delta
-4 \epsilon^m (1 + \rho_m)^m L^{(m)}_1 / \delta$
to $\theta_{m+1}$ on the domain 
$|{\rm Im} \theta_{m+1}| \le \rho_m - 2 \delta
-3 \epsilon^m (1 + \rho_m)^m L^{(m)}_1 / \delta$.
So on the domain $|J_{m+1} - J_{m+1~0}| \le \rho_m - \delta$,
$|{\rm Im} \theta_{m+1}| \le \rho_m - 2 \delta
-3 \epsilon^m (1 + \rho_m)^m L^{(m)}_1 / \delta$,
the inverse mapping $T^{-1}_{J_{m+1}}$ is defined and diffeomorphic
and $|{\rm Im} T^{-1}_{J_{m+1}} \theta_{m+1}| = |{\rm Im} \theta_m|
\le \rho_m - 2 \delta - 4 \epsilon^m (1 + \rho_m)^m 
L^{(m)}_1 / \delta$.
From now on, we assume that $\epsilon$ satisfies 
\begin{equation}
 3 \epsilon^m (1 + \rho_m)^m {L^{(m)}_1 \over \delta}
 \le \delta.
\end{equation}
Then the canonical transformation 
$(J_{m+1}, \theta_{m+1}) \rightarrow (J_m, \theta_m)$
induced by the generating function $S_m(J_{m+1}, \theta_m, t)$
is defined and is diffeomorphic
on the domain $|J_{m+1} - J_{m+1~0}| \le \rho_m -\delta$, 
$|{\rm Im} \theta_m| \le \rho_m-3 \delta$ and
$|{\rm Im} T^{-1}_{J_{m+1}} \theta_{m+1}| = |{\rm Im} \theta_m|
\le \rho_m - 3 \delta + {\delta / 3} \le \rho_m - 2 \delta$.

Next on the domain $|J_{m+1} - J_{m+1~0}| \le \rho_m-2 \delta$, 
$|{\rm Im} \theta_m| \le \rho_m-3 \delta$,
we evaluate the residual part $R$ defined as 
\begin{equation}
 R = R_1 + R_2 + R_3 +
 {\partial S_m (J_{m+1}, \theta_m, t)
  \over \partial t }.
\end{equation}
$R_1$ can be rewritten as
\begin{eqnarray}
 R_1 &=& \int^{J_m}_{J_{m+1}} \mu \cdot dJ
  {1 \over \epsilon} 
       \{
  {1 \over (\mu \cdot J)^{1 / 2}} -
  {1 \over (\mu \cdot J_{m+1})^{1 / 2}}
       \}
\nonumber \\
  &=& - \int^{J_m}_{J_{m+1}} \mu \cdot dJ
  {1 \over 2 \epsilon} 
      \bigr\{
  \int_{J_{m+1}} \mu \cdot dJ
  {1 \over (\mu \cdot J)^{3 / 2}}
       \bigl\}.
\end{eqnarray}
We evaluate
\begin{eqnarray}
 |t^{2 m} R_1| &\le& \epsilon^{2 m -1} \mu^2 N^2 
 (1 + \rho_m)^{2 m} {L^{(m)2}_1 \over \delta^2} 
 {1 \over \sigma^{3 / 2}_m}
\nonumber \\
 &\le&  \epsilon^m \mu^2 N^2 
 (1 + \rho_m)^{2 m} {L^{(m)2}_1 \over \delta^2} 
 {1 \over \sigma^{3 / 2}_m}.
\end{eqnarray}
Therefore we obtain
\begin{equation}
 |t^{m + 1} R_1| \le \epsilon^m \mu^2 N^2 
 (1 + \rho_m)^{3 m - 1} {L^{(m)2}_1 \over \delta^2} 
 {1 \over \sigma^{3 / 2}_m}.
\end{equation}

As for $R_2$, we obtain
\begin{eqnarray}
 |t^{m+2} R_2| &=& |t^{m + 2} \int^{J_m}_{J_{m+1}}
 {\partial A_m (J_m, t) \over \partial J_m} dJ_m |
\nonumber \\
 &\le& 
 \sup\{ |t^2 {\partial A_m (J_m, t) \over \partial J_m}| \}
 | t^m 
 {\partial S_m (J_{m+1}, \theta_m, t) \over \partial \theta_m} |
\nonumber \\
 &\le&
 \epsilon^m {M^{(m)}_1 L^{(m)}_1 \over 6 \delta^2}.
\end{eqnarray}
Therefore we obtain
\begin{equation}
 |t^{m + 1} R_2| \le  
 \epsilon^m (1 + \rho_m){M^{(m)}_1 L^{(m)}_1 \over 6 \delta^2}.
\end{equation}

As for $R_3$, we obtain
\begin{eqnarray}
 |t^{2 m} R_3| &=& |t^{2 m} \int^{J_m}_{J_{m+1}}
 {\partial B_m (J_m, \theta_m, t) \over \partial J_m} dJ_m |
\nonumber \\
 &\le& 
 \sup\{ |t^m 
        {\partial B_m (J_m, \theta_m, t) \over \partial J_m}| \}
 | t^m
 {\partial S_m (J_{m+1}, \theta_m, t) \over \partial \theta_m} |
\nonumber \\
 &\le&
 \epsilon^{2 m-1} {M^{(m)}_2 L^{(m)}_1 \over 6 \delta^2}
\nonumber \\
 &\le&
 \epsilon^m {M^{(m)}_2 L^{(m)}_1 \over 6 \delta^2}
\end{eqnarray}
Therefore we obtain
\begin{equation}
 |t^{m + 1} R_3| \le  
 \epsilon^m (1 + \rho_m)^{m-1}
 {M^{(m)}_2 L^{(m)}_1 \over 6 \delta^2}.
\end{equation}

Also we obtain
\begin{equation}
 |t^{m + 1} {\partial S_m \over \partial t}| \le
 \epsilon^m [
 (m-1) L^{(m)}_1 + (1 + \rho_m ) {L^{(m)}_1 \over \delta} ].
\end{equation}

So $R$ is evaluated as 
\begin{equation}
 |t^{m+1} R| \le \epsilon^m M,
\end{equation}
where
\begin{eqnarray}
 M &=& \mu^2 N^2 (1 + \rho_m)^{3 m - 1} {L^{(m)2}_1 \over \delta^2}
 {1 \over \sigma^{3 / 2}_m} +
 (1 + \rho_m) { M^{(m)}_1 L^{(m)}_1 \over 6 \delta^2 }+
 (1 + \rho_m)^{m - 1}  
 { M^{(m)}_2 L^{(m)}_1 \over 6 \delta^2 }
\nonumber \\
&& + (m-1) L^{(m)}_1 +
   (1 + \rho_m ) {L^{(m)}_1 \over \delta},
\end{eqnarray}
on the domain 
$(J_{m+1}, \theta_{m+1}, {1 / t}) \in D_{m+1}(\rho_m - 2 \delta)$.
We decompose $R$ into the fourier components as
\begin{eqnarray}
 R(J_{m+1}, \theta_{m+1}, t) = \sum_k R_k (J_{m+1}, t) 
 e^{i k \theta_{m+1}}.
\end{eqnarray}
As for $A_{m+1}$, $B_{m+1}$ defined as 
\begin{eqnarray}
 &&
 A_{m+1} (J_{m+1}, t) = A_m (J_{m+1}, t)+R_0 (J_{m+1}, t),\\
 &&
 B_{m+1} (J_{m+1}, \theta_{m+1}, t) = 
 \sum_{k \ne 0} R_k (J_{m+1}, t) e^{i k \theta_{m+1}},
\end{eqnarray}
we obtain
\begin{eqnarray}
 &&
 |t^2 A_{m+1}| \le M^{(m+1)}_1,\\
 &&
 |t^{m+1} B_{m+1}| \le \epsilon^m M^{(m+1)}_2,
\end{eqnarray}
where
\begin{eqnarray}
 &&
 M^{(m+1)}_1 =  M^{(m)}_1 + (1 + \rho_m)^{m-2} \epsilon^m M,
\\
 &&
 M^{(m+1)}_2 = 2 M
\end{eqnarray}
on the domain 
$(J_{m+1}, \theta_{m+1}, {1 / t}) \in D_{m+1} (\rho_m - 2 \delta)$.
We put $\rho_m = \rho_{m-1} - 2 \delta$.
We complete the proof.

{\it Proof End}

The following proposition is used in evaluating the difference
between the original perturbation variables and the transformed 
perturbation variables.
Since as shown later, the action variable perturbations and the angle
variable perturbations are bounded by $t^0$, $t$ respectively,
the difference between the original perturbation variables and 
the transformed perturbation variables is bounded by some positive 
constant.

{\bf Proposition2}

Suppose that the generating functions $S_m$ are analytic and 
the inequalities
\begin{eqnarray}
 &&
 |t^m S_m| \le \epsilon^m L_1^{(m)},
\\
 &&
 |t^m ({\partial \over \partial J_{m+1}})^{\alpha_1}
      ({\partial \over \partial \theta_m })^{\alpha_2} S_m| \le 
 \epsilon^m L_2^{(m)}~~~(\alpha_1 + \alpha_2 \le 2)
\end{eqnarray}
holds for some positive $L_1^{(m)}$, $L_2^{(m)}$.

As for $m \ge 2$, the inequalities 
\begin{eqnarray}
 &&
 |\delta J^i_1(t)- \delta J^i_m(t)| \le 
   {\epsilon \over t} C^{(m)}_{11} |\delta J_m(t)|
 + {\epsilon \over t} C^{(m)}_{12} |\delta \theta_m(t)|,
\label {diffdelJ}
\\
 &&
 |\delta \theta^i_1(t)- \delta \theta^i_m(t)| \le 
   {\epsilon \over t} C^{(m)}_{21} |\delta J_m(t)|
 + {\epsilon \over t} C^{(m)}_{22} |\delta \theta_m(t)|
\label {diffdelth}
\end{eqnarray}
holds for some positive
$C^{(m)}_{11}$, $C^{(m)}_{12}$, $C^{(m)}_{21}$, $C^{(m)}_{22}$,
where
\begin{eqnarray}
&&
 |\delta J_m(t)| \equiv \max_{1 \le i \le N} |\delta J^i_m(t)|,
\\
&&
 |\delta \theta_m(t)| \equiv 
 \max_{1 \le i \le N} |\delta \theta^i_m(t)|.
\end{eqnarray}

{\it Proof}

By taking the variation of
\begin{equation}
 J^i_m-J^i_{m+1} = {\partial S_m \over \partial \theta^i_m},
\quad \quad
 \theta^i_{m+1} - \theta^i_m ={\partial S_m \over \partial J^i_{m+1} },
\end{equation}
we obtain
\begin{eqnarray}
 &&
 |\delta J^i_m - \delta J^i_{m+1}| \le {\epsilon^m \over t^m}
   N L^{(m)}_2 ( |\delta J_{m+1}|+|\delta \theta_m| ),
\label {diff1}
\\
 &&
 |\delta \theta^i_{m+1} - \delta \theta^i_m| \le 
  {\epsilon^m \over t^m }
   N L^{(m)}_2 ( |\delta J_{m+1}|+|\delta \theta_m| ).
\label {diff2}
\end{eqnarray}
As for $|\delta \theta^i_m|$, we obtain
\begin{eqnarray}
 |\delta \theta^i_m| &\le& 
 |\delta \theta^i_m - \delta \theta^i_{m+1}| + |\delta \theta^i_{m+1}|
\nonumber \\
 &\le&
 {\epsilon^m \over t^m}
 N L^{(m)}_2 ( |\delta J_{m+1}|+|\delta \theta_m| )
 +|\delta \theta_{m + 1}|,
\end{eqnarray}
which yield
\begin{equation}
 |\delta \theta_m| \le {\epsilon^m \over t^m} N
 {L^{(m)}_2}' |\delta J_{m+1}| 
 + (1+ \epsilon^{m-1} N {L_2^{(m)}}'')|\delta \theta_{m+1}|.
\end{equation}
By using the above inequality, 
the inequalities (\ref {diff1}), (\ref {diff2}) are rewritten
as
\begin{eqnarray}
 &&
 |\delta J^i_m - \delta J^i_{m+1}| \le {\epsilon^m \over t^m}
   N {L^{(m)}_2}''' ( |\delta J_{m+1}|+|\delta \theta_{m+1}| ),
\\
 &&
 |\delta \theta^i_{m+1} - \delta \theta^i_m| \le 
  {\epsilon^m \over t^m}
   N {L^{(m)}_2}''' ( |\delta J_{m+1}|+|\delta \theta_{m+1}| ).
\end{eqnarray}
By using the above inequalities, by the induction, it is shown that 
for an arbitrary natural number $m \ge 2$, the inequalities 
\begin{eqnarray}
 &&
 |\delta J^i_1(t)- \delta J^i_m(t)| \le 
   {\epsilon \over t} C^{(m)}_{11} |\delta J_m(t)|
 + {\epsilon \over t} C^{(m)}_{12} |\delta \theta_m(t)|,
\\
 &&
 |\delta \theta^i_1(t)- \delta \theta^i_m(t)| \le 
   {\epsilon \over t} C^{(m)}_{21} |\delta J_m(t)|
 + {\epsilon \over t} C^{(m)}_{22} |\delta \theta_m(t)|
\end{eqnarray}
holds for some positive
$C^{(m)}_{11}$, $C^{(m)}_{12}$, $C^{(m)}_{21}$, $C^{(m)}_{22}$.

{\it Proof End}

\section{ Evolution of the exactly homogeneous universe }

In this section,
we investigate the evolutionary behavior of the exactly homogeneous 
universe dominated by scalar fields.
In terms of action angle variables, the evolution of the 
scalar fields is characterized by the perpetual stability
of the action variables which we formulate below.

{\bf Proposition3 (Perpetual stability of the action variables)}

Suppose that the hamiltonian 
\begin{equation}
 H = {2 \over \epsilon} 
            \Bigl(
    {1 \over 2} {1 \over t^2} \sum_i P_{\phi_i}^2 
    + {1 \over 2} t^2 \sum_i \mu_i^2 \Phi_i^2 
    + \lambda t^2 U_{\rm int}(\Phi)
            \Bigr)^{1/2}
\end{equation}
which satisfy the following conditions;

(i) For some positive $C$ and $d$, the inequality
\begin{equation}
 |k \cdot \mu| \ge C {1 \over |k|^{d} }
\end{equation}
holds for every integral-valued vector $k=(k_1, \cdot \cdot \cdot, k_N)$.

In terms of action angle variables introduced by
\begin{eqnarray}
 \Phi_i &=& \sqrt{ {2 \over \mu_i} } \sqrt{J_i} {1 \over t} 
 \cos{\theta_i},\\
 P_{\phi_i} &=& - \sqrt{ 2 \mu_i } \sqrt{J_i} t \sin{\theta_i},
\end{eqnarray}

(ii)$U_{\rm int}(\Phi)$ is a sum of monomials in terms of $\Phi$
of power not smaller than $3$.
The inequality 
\begin{equation}
 | \lambda t^2 U_{\rm int}(\Phi) | \le \epsilon M
\end{equation}
holds for some positive $M$ on the domain $(J, \theta,  {1 / t}) \in 
D(\rho)$

(iii) The inequality 
\begin{equation}
  | {1 \over 2} {1 \over t^2} \sum_i P_{\phi_i}^2 
    + {1 \over 2} t^2 \sum_i \mu_i^2 \Phi_i^2 | =
  |\mu \cdot J|
  > \sigma
\end{equation}
holds for some positive $\sigma$ on the domain 
$(J, {1 / t}) \in D'(\rho)$.

There exists some positive $\epsilon_0$ such that 
for an arbitrary $\epsilon$ satisfying
\begin{equation}
 0 < \epsilon < \epsilon_0,
\end{equation}
the inequality 
\begin{equation}
 |J - J_0| \le \epsilon C
\end{equation}
holds for some positive $C$.

{\it Proof}

We consider the sequence of canonical transformations defined 
in the {\bf Proposition1}.
\begin{equation}
H^{(1)} (J_1, \theta_1, t)
 \stackrel{S_1}{\rightarrow}
H^{(2)} (J_2, \theta_2, t).
\end{equation}
The transformed hamiltonian is given by
\begin{equation}
 H^{(2)} = {2 \over \epsilon}(\mu \cdot J_2)^{1 / 2} 
     + A_2 (J_2, t)
     + B_2 (J_2, \theta_2, t),
\end{equation}
where 
\begin{equation}
 {1 \over (2 \pi)^N} \int_0^{2 \pi} \cdot \cdot \cdot \int_0^{2 \pi}
d^N \theta_3 
 t^2 B_2 (J_2, \theta_2, t) = 0.
\end{equation}
The inequality 
\begin{equation} 
 |t^2 A_2| \le M^{(2)}_1, ~~~|t^2 B_2| \le \epsilon M^{(2)}_2
\end{equation}
is satisfied on the domain $D_2(\rho_2)$.
The equation of $J_2$ is 
\begin{equation}
 {d J^i_2 \over d t} = - {\partial B_2 \over \partial \theta^i_2}.
\end{equation}
For an arbitrary positive $\delta$,
on the domain $D(\rho_2 - \delta)$ 
\begin{eqnarray}
 | J^i_2 (t) - J^i_{2~0} | &\le& \int_1 dt 
 | {\partial B_2 \over \partial \theta^i_2} |
\nonumber \\
 &\le& \int_1 dt {\epsilon \over t^2} { M^{(2)}_2 \over \delta} 
\nonumber \\
 &\le& \epsilon { M^{(2)}_2 \over \delta}.
\end{eqnarray}
for an arbitrary $t \ge 1$.

For an arbitrary positive $\epsilon$ satisfying
\begin{equation}
 \sqrt{N} \epsilon { M^{(2)}_2 \over \delta} \le \rho_2 - \delta
\end{equation}
the inequality
\begin{equation}
 | J_2 - J_{20} | \le \sqrt{N} \epsilon {M^{(2)}_2 \over \delta}. 
\end{equation}
holds.

We complete the proof.

{\it Proof End}

\section{Evolution of Cosmological Perturbations 
in the Long Wavelength Limit}

We consider the evolution of the cosmological perturbations 
in the long wavelength limit 
since it approximates the evolution of the cosmological perturbations 
on superhorizon scales well \cite{Kodama.H&Hamazaki1996}.
In this section, by using the method of the long wavelength limit,
and the canonical perturbation theory, we investigated 
the evolutionary behavior of the cosmological perturbations, 
in particular paying attention to that of the Bardeen parameter $Z$.
We show that for the adiabatic growing mode $Z$ stays constant,
and for all the other modes $Z$ decays.
Since the Bardeen parameter becomes singular probabilistically
when the masses of the scalar fields are incommensurable,
we discuss the distribution of the singularities of $Z$ in the time
axis in terms of the ergodic theory.

On superhorizon scales, the Bardeen parameter defined by
\begin{equation}
 Z = {\cal R} - {a H \over k} \sigma_g,
\end{equation}
is conserved in a good accuracy when the equation of state is regular 
and the energy of the universe is dominated by a single component.
On the other hand, in the universe dominated 
by multi-component scalar fields, the Bardeen parameter is rewritten as 
\begin{equation}
 Z = - H {\dot{\phi}\cdot Y \over (\dot{\phi})^2},
\end{equation}
whose conservation is not self-evident.
By using Eq.(\ref{k0limitWronskian}),
we can see that
the contribution from the second term in the right hand side of 
Eq.(\ref{fieldperturbationformula}) is given by
\begin{equation}
 Z \propto \int dt {1 \over a^3}.
\end{equation}
So we concentrate the contribution from the first term in the right hand side of 
Eq.(\ref{fieldperturbationformula}) henceforth.
\begin{equation}
 Z = - H {\dot{\phi}\cdot \delta \phi \over (\dot{\phi})^2}.
\label{Bardeenhomogeneouscontribution}
\end{equation}

The contribution to the Bardeen parameter from the exactly homogeneous 
perturbation, Eq.(\ref{Bardeenhomogeneouscontribution}), is rewritten 
in terms of the action angle variables as
\begin{eqnarray}
 Z &=& {2 \over 3} {\epsilon \over t} (\mu \cdot J + K)^{1/2} 
       {1 \over \sum_i \mu_i J^i (1 - \cos{2 \theta^i})}
\nonumber \\
   && \sum_i \Bigl(
        {1 \over 2} \delta J^i \sin{2 \theta^i} 
         - J^i (1 - \cos{2 \theta^i}) \delta \theta^i
              \Bigr). 
\label{Bardeenactionangle}   
\end{eqnarray}

In order to understand the evolutionary behavior of the perturbation variables
such as $Z$, we have to investigate that of $\delta \theta^i$, $\delta J^i$.  
Since the oscillating part is troublesome for investigating the differential
equations of perturbations, we eliminate the fast phases by the canonical 
transformations.
We consider the sequence of canonical transformations defined 
in the {\bf Propositions1}. 
\begin{equation}
                               H^{(1)} (J_1, \theta_1, t)
  \stackrel{S_1}{\rightarrow}  H^{(2)} (J_2, \theta_2, t)
  \stackrel{S_2}{\rightarrow}  H^{(3)}  (J_3, \theta_3, t).
\end{equation}
The transformed hamiltonian is given by
\begin{equation}
 H^{(3)} = {2 \over \epsilon}(\mu \cdot J_3)^{1 / 2} + A_3 (J_3, t)
     + B_3 (J_3, \theta_3, t),
\end{equation}
where 
\begin{equation} 
 |t^2 A_3| \le M^{(3)}_1, ~~~|t^3 B_3| \le \epsilon^2 M^{(3)}_2
\end{equation}
is satisfied on the domain $D_3(\rho_3)$.
The variational equation of motion changed into the integral form is 
\begin{eqnarray}
 &&
 \delta J^i_3 = \delta J^i_{30} - \int_1 dt
        \bigl[
 {\partial^2 B_3 \over \partial \theta^i_3 \partial \theta_3}
 \delta \theta_3
 + {\partial^2 B_3 \over \partial \theta^i_3 \partial J_3}
 \delta J_3
        \bigr],
\\
 &&
 {\delta \theta^i_3 \over t} =
 {\delta \theta^i_{30} \over t} +
 {1 \over t}{\mu_i \over \epsilon}
 \int_1 dt \delta {1 \over (\mu \cdot J_3)^{1 / 2}} 
\nonumber \\
 && \quad \quad \quad + {1 \over t}
 \int_1 dt  
 {\partial^2 A_3 \over \partial J^i_3 \partial J_3}
 \delta J_3 +
 {1 \over t}
 \int_1 dt  
         \bigl[
 {\partial^2 B_3 \over \partial J^i_3 \partial \theta_3}
 \delta \theta_3
 + {\partial^2 B_3 \over \partial J^i_3 \partial J_3}
 \delta J_3
         \bigr].
\label {deltatheta4}
\end{eqnarray}

For an arbitrary positive $\delta$, 
on the domain $D_3(\rho_3 - \delta)$, we obtain
\begin{eqnarray}
 &&
 |\delta J^i_3 (t)| \le |\delta J^i_{30}|
 + {2 \over \delta^2} \epsilon^2 M^{(3)}_2 
   \bigl\{
 |{\delta \theta_3 \over t}|_m (t)
 + {1 \over 2} |\delta J_3|_m (t)
   \bigr\},
\\
 &&
 |{\delta \theta^i_3 (t) \over t}| \le
 |\delta \theta^i_{30}| +
 {\mu_i \over \epsilon} {1 \over 2}
 {N \mu \over \sigma^{3 / 2}_3}
 |\delta J_3|_m (t)
 + {1 \over t} {2 \over \delta^2}
 M^{(3)}_1 |\delta J_3|_m (t)
\nonumber \\
 &&
 \quad \quad \quad \quad
 + {1 \over t} {2 \over \delta^2}
 \epsilon^2 M_2^{(3)} 
   \bigl\{
 |{\delta \theta_3 \over t}|_m (t)
 + {1 \over 2} |\delta J_3|_m (t)
   \bigr\},
\end{eqnarray}
where
\begin{equation}
 |A|_m (t) = \sup_{1 \le s \le t} {|A(s)|}
\end{equation}
They yield
\begin{eqnarray}
 &&
 |\delta J_3|_m (t) \le (1 + \epsilon C_1) |\delta J_{30}|
                        + \epsilon^2 C_2 |\delta \theta_{30}|,
\label{resultJ4}
\\
 &&
 |{\delta \theta_3 \over t}|_m (t) \le 
 {1 \over \epsilon} C_3 |\delta J_{30}| +
 (1 + \epsilon C_4) |\delta \theta_{30}|,
\label{resulttheta4}
\end{eqnarray}
for some positive $C_1$, $\cdot \cdot \cdot$ $C_4$.
In particular, when Eq.(\ref{deltatheta4}) is written as 
\begin{eqnarray}
 &&
 \delta \theta^i_3 = \delta \theta^i_{30} 
   + {\mu_i \over \epsilon} (t - 1) V_3 (t)
   + W^i_3 (t),
\\
 && \quad \quad
 V_3 (t) = {1 \over (t-1)} \int_1 dt 
           \delta{1 \over (\mu \cdot J_3)^{1 / 2}},\\
 && \quad \quad
 W^i_3 (t) = \int_1 dt  
 {\partial^2 A_3 \over \partial J^i_3 \partial J_3}
 \delta J_3 + \int_1 dt  
         \bigl[
 {\partial^2 B_3 \over \partial J^i_3 \partial \theta_3}
 \delta \theta_3
 + {\partial^2 B_3 \over \partial J^i_3 \partial J_3}
 \delta J_3
         \bigr],
\end{eqnarray}
from the inequalities (\ref {resultJ4}), (\ref {resulttheta4}), 
we obtain 
\begin{eqnarray}
 &&
 |V_3 (t) - \delta {1 \over (\mu \cdot J_{30})^{1 / 2}}| \le 
 \epsilon ( C_5 |\delta J_{30}| + \epsilon C_6 |\delta \theta_{30}| ),
\label {V4}\\
 &&
 |W^i_3 (t)| \le C_7 |\delta J_{30}| + \epsilon^2 C_8 |\delta \theta_{30}|.
\label {W4}
\end{eqnarray}

Next we have to know the difference between the original variables 
$\delta \theta_1$, $\delta J_1$ and 
the transformed variables $\delta \theta_3$, $\delta J_3$,
since we want to know the information of the original variables. 
As for the initial values between  
of $\delta \theta_1$, $\delta J_1$ and of $\delta \theta_m$, $\delta J_m$,
the inequalities
\begin{eqnarray}
 &&
 |\delta J^i_{10}- \delta J^i_{m0}| \le 
   \epsilon C'^{(m)}_{~11} |\delta J_{10}|
 + \epsilon C'^{(m)}_{~12} |\delta \theta_{10}|,
\\
 &&
 |\delta \theta^i_{10}- \delta \theta^i_{m0}| \le 
   \epsilon C'^{(m)}_{~21} |\delta J_{10}|
 + \epsilon C'^{(m)}_{~22} |\delta \theta_{10}|.
\end{eqnarray}
and the inequalities 
\begin{eqnarray}
&&
 |\delta J_{m0}| \le (1 + \epsilon C^{(m)}_{33}) |\delta J_{10}|
                    + \epsilon C^{(m)}_{34} |\delta \theta_{10}|,
\label{deltaJm0}
\\
&&
 |\delta \theta_{m0}| \le \epsilon C^{(m)}_{43} |\delta J_{10}|
                    + (1 + \epsilon C^{(m)}_{44}) |\delta \theta_{10}|,
\label{deltathetam0}
\end{eqnarray}
hold. 
Using (\ref {deltaJm0}), (\ref {deltathetam0}) to 
(\ref {resultJ4}), (\ref {resulttheta4})
we obtain
\begin{eqnarray}
 &&
 |\delta J_3|_m (t) \le (1 + \epsilon C^{(4)}_1 ) |\delta J_{10}|
                        + \epsilon C^{(4)}_2 |\delta \theta_{10}|,
\label{deltaJ4t}
\\
 &&
 |{\delta \theta_3 \over t}|_m (t) \le 
                       {1 \over \epsilon} C^{(4)}_3 |\delta J_{10}|
                    + C^{(4)}_4 |\delta \theta_{10}|.
\label{deltatheta4t}
\end{eqnarray}
By using (\ref {deltaJ4t}), (\ref {deltatheta4t}) to 
(\ref {diffdelJ}), (\ref {diffdelth}), we obtain
\begin{eqnarray}
 &&
 |\delta J^i_1 (t) - \delta J^i_3 (t)| \le 
  C_1 |\delta J_{10}| + \epsilon C_2 |\delta \theta_{10}|,
\label {diffJ}\\
 &&
 |\delta \theta^i_1 (t) - \delta \theta^i_3 (t)| \le 
  C_3 |\delta J_{10}| + \epsilon C_4 |\delta \theta_{10}|.
\label {difftheta}
\end{eqnarray}
From (\ref {deltaJ4t}), (\ref {diffJ}), we obtain
\begin{equation}
 |\delta J^i_1 (t)| \le 
 C_5 |\delta J_{10}| + \epsilon C_6 |\delta \theta_{10}|.
\end{equation}
This yields the evaluation (\ref {fluctuationdeltaJ}).

On the other hand, from (\ref {V4}), (\ref {W4}), (\ref {deltaJm0}), 
(\ref {deltathetam0}), (\ref {difftheta}), 
we obtain the evaluation (\ref {fluctuationdeltaJ}), 
(\ref {fluctuationdeltatheta1}), $\cdot \cdot \cdot$ 
(\ref {fluctuationdeltatheta3}).

{\bf Proposition4 (Evaluation of the perturbation variables)}

Suppose that the hamiltonian 
\begin{equation}
 H = {2 \over \epsilon} 
            \Bigl(
    {1 \over 2} {1 \over t^2} \sum_i P_{\phi_i}^2 
    + {1 \over 2} t^2 \sum_i \mu_i^2 \Phi_i^2 
    + \lambda t^2 U_{\rm int}(\Phi)
            \Bigr)^{1/2}
\end{equation}
which satisfy the following conditions;

(i) For some positive $C$ and $d$, the inequality
\begin{equation}
 |k \cdot \mu| \ge C {1 \over |k|^{d} }
\end{equation}
holds for every integral-valued vector $k=(k_1, \cdot \cdot \cdot, k_N)$.

In terms of action angle variables introduced by
\begin{eqnarray}
 \Phi_i &=& \sqrt{ {2 \over \mu_i} } \sqrt{J_i} {1 \over t} 
 \cos{\theta_i},\\
 P_{\phi_i} &=& - \sqrt{ 2 \mu_i } \sqrt{J_i} t \sin{\theta_i},
\end{eqnarray}

(ii) $U_{\rm int}(\Phi)$ is a sum of monomials in terms of $\Phi$
of power not smaller than $3$.
The inequality
\begin{equation}
 |\lambda t^2 U_{\rm int}(\Phi)| \le \epsilon M
\end{equation}
holds for some positive $M$ 
on the domain $(J, \theta, 1 / t) \in D(\rho)$.

(iii) The inequality 
\begin{equation}
  | {1 \over 2} {1 \over t^2} \sum_i P_{\phi_i}^2 
    + {1 \over 2} t^2 \sum_i \mu_i^2 \Phi_i^2 |
  = |\mu \cdot J|
  > \sigma
\end{equation}
holds for some positive $\sigma$ on the domain 
$(J, 1 / t) \in D'(\rho)$

There exists some positive $\epsilon_0$ such that; 

for an arbitrary $\epsilon$ satisfying
\begin{equation}
 0 < \epsilon < \epsilon_0,
\end{equation}
the following inequalities holds.

As for $\delta J^i(t)$, the inequality 
\begin{equation}
 |\delta J^i (t) - \delta J^i_0| \le C_1 |\delta J_0| 
                     + \epsilon C_2 |\delta \theta_0|
\label {fluctuationdeltaJ}
\end{equation}
holds for some positive $C_1$, $C_2$.

When $\delta \theta^i(t)$ is decomposed as 
\begin{equation}
 \delta \theta^i(t) = \delta \theta^i_0 +
 {\mu_i \over \epsilon} (t-1) V(t) + W^i (t).
\label {fluctuationdeltatheta1}
\end{equation}

(i) $V(t)$ is $i$-indendent and the inequality
\begin{equation}
 |V(t) - \delta ( {1 \over \mu \cdot J_0} )| \le \epsilon 
 ( C_3 |\delta J_0| + \epsilon C_4 |\delta \theta_0| ) 
\label {fluctuationdeltatheta2}
\end{equation}
holds for some positive $C_3$ and $C_4$.

(ii) $W^i (t)$ is $i$-dependent in general and the inequality
\begin{equation}
 |W^i (t)| \le C_5 |\delta J_0| + \epsilon C_6 |\delta \theta_0|
\label {fluctuationdeltatheta3}
\end{equation} 
holds for some positive $C_5$ and $C_6$.

The decomposition of $\delta \theta^i (t)$ into $V(t)$ and 
$W^i (t)$ as in (\ref {fluctuationdeltatheta1}) is not 
unique.
For the purpose explained below, it may be rather convenient 
to write in the next form.

{\bf Proposition4'}

Under the same assumption presented in the above proposition,
for $t \ge 2$, there exists a function $V(t)$ such that
\begin{eqnarray}
 &&
 \delta \theta^i(t) = \delta \theta^i_0 +
 {\mu_i \over \epsilon} (t-1) V(t) + W^i (t),
\\
 &&
 |V(t) - V(t=\infty)| \le {\epsilon^2 \over t} 
 ( C_1 |\delta J_0| + \epsilon C_2 |\delta \theta_0| ),
\\ 
 &&
 |W^i (t)| \le C_3 |\delta J_0| + \epsilon C_4 |\delta \theta_0|,
\end{eqnarray}
for some positive $C_1$, $\cdot \cdot \cdot$, $C_4$.

{\it Proof}

We consider the transformed hamiltonian given by
\begin{equation}
 H^{(4)} = {2 \over \epsilon}(\mu \cdot J_4)^{1 / 2} + A_4 (J_4, t)
     + B_4 (J_4, \theta_4, t),
\end{equation}
where 
\begin{equation} 
 |t^2 A_4| \le M^{(4)}_1, ~~~|t^4 B_4| \le \epsilon^3 M^{(4)}_2
\end{equation}
is satisfied on the domain $D_4(\rho_4)$.
As for $J^i_4 (t)$ and $\delta J^i_4 (t)$, we obtain inequalities
\begin{eqnarray}
 |J^i_4 (t) - J^i_{4~\infty}| &\le& 
  \int^{\infty}_t dt
        \bigl|
 {\partial B_4 \over \partial \theta^i_4}
        \bigl|
\nonumber \\
 &\le&
 \epsilon^3 C_1 {1 \over t^3},
\label {inftyJi}
\\
 |\delta J^i_4 (t) - \delta J^i_{4~\infty}| &\le& 
  \int^{\infty}_t dt
        \bigl|
 {\partial^2 B_4 \over \partial \theta^i_4 \partial \theta_4}
 \delta \theta_4
 + {\partial^2 B_4 \over \partial \theta^i_4 \partial J_4}
 \delta J_4
        \bigr|
\nonumber \\
 &\le&
 {\epsilon^2 \over t^2} 
 (C_2 |\delta J_0| + \epsilon C_3 |\delta \theta_0|),
\label {inftydelJi}
\end{eqnarray}
for some positive $C_1$, $C_2$, $C_3$.
As $V(t)$ in the proposition, we adopt
\begin{equation}
 V(t) = {1 \over t-1} \int^t_1 dt 
 \delta {1 \over (\mu \cdot J_4)^{1 / 2}}.
\label {inftydeltaJi}
\end{equation}
By using (\ref {inftyJi}), (\ref {inftydelJi}), (\ref {inftydeltaJi}), 
for $t_1, t_2 \ge 2$, we obtain
\begin{eqnarray}
&&
 |V(t_1) - V(t_2)| = 
   \bigl|
 {1 \over t_1 - 1} \int^{t_1}_{t_2} dt 
 \delta {1 \over (\mu \cdot J_4)^{1 / 2}}
 + {(t_2 - t_1) \over (t_1 - 1) (t_2 - 1)}
 \int^{t_2}_1 dt 
 \delta {1 \over (\mu \cdot J_4)^{1 / 2}}
   \bigr|
\nonumber \\
 && \le
    \bigl|
{-1 \over t_1 -1} \int^{t_2}_{t_1} dt
(-{1 \over 2}) {1 \over (\mu \cdot J_4)^{3 / 2}}
\mu \cdot \delta J_4
-
{-1 \over t_1 -1} \int^{t_2}_{t_1} dt
(-{1 \over 2}) {1 \over (\mu \cdot J_{4~{\infty}})^{3 / 2}}
\mu \cdot \delta J_{4~{\infty}}
    \bigr|
\nonumber \\
 && +
    \bigl|
 {t_2 - t_1 \over (t_1 - 1)(t_2 - 1)} \int^{t_2}_1 dt
(-{1 \over 2}) {1 \over (\mu \cdot J_4)^{3 / 2}}
\mu \cdot \delta J_4
\nonumber \\
 && -
 {t_2 - t_1 \over (t_1 - 1)(t_2 - 1)} \int^{t_2}_1 dt
(-{1 \over 2}) {1 \over (\mu \cdot J_{4~{\infty}})^{3 / 2}}
\mu \cdot \delta J_{4~{\infty}} 
    \bigr|
\nonumber \\
 && \le
    \bigl|
 {1 \over t_1} - {1 \over t_2}
    \bigr|
 \epsilon^2 
 (C_4 |\delta J_0| + \epsilon C_5 |\delta \theta_0|),
\end{eqnarray}
which yields 
\begin{equation}
 |V(t) - V(t=\infty)| \le {\epsilon^2 \over t}
 (C_4 |\delta J_0| + \epsilon C_5 |\delta \theta_0|).
\end{equation}
We complete the proof.

{\it Proof End}

Using the above results to Eq.(\ref{Bardeenactionangle})
we can see that for sufficiently large $t$, 
\begin{equation}
 Z \rightarrow - {2 \over 3} (\mu \cdot J + K)^{1 / 2} 
    V(t=\infty).
\label {growingterm}
\end{equation}
Therefore for the mode satisfying $V(t=\infty) \ne 0$
which corresponds to the adiabatic growing mode, Z stays constant,
and for other $(2 N - 1)$ modes satisfying $V(t=\infty) = 0$, 
Z decays at most in proportion to $1 / t$. 
Then the uniqueness of the growing mode are derived.

Because of the factor as
\begin{equation}
 {1 \over \sum_i \mu_i J^i (1 - \cos 2 \theta^i)},
\end{equation}
all the terms in Eq.(\ref{Bardeenactionangle}) 
except (\ref {growingterm}) become boundlessly large when the phase 
flow ${\bf \theta} (t)$ approaches $\theta^i = 0$ or $\pi$ 
$(i=1, \cdot \cdot \cdot, N)$ boundlessly, 
although the numerator are not taken into account approaching zero.
Does our system spend time near $\theta^i = 0$ or $\pi$ 
$(i=1, \cdot \cdot \cdot, N)$ where the Bardeen parameter becomes 
large?
The answer is as follows.
Our track approaches the singular point boundlessly 
infinite times. 
The distribution of the singularities in the time axis is 
probabilistic and random.
This problem can be regarded in the viewpoint 
of the ergodic theory.
We formulate this as follows.

{\bf Proposition5}

Under the condition

(i)' If the equality $\mu \cdot k = 0$ holds for some integral-valued
vector $k = (k^1, \cdot \cdot \cdot ,k^N)$, $k=0$ holds.

and the assumptions (ii), (iii) in the above proposition, 

there exists some positive $\epsilon_0$ such that; 

for an arbitrary $\epsilon$ satisfying
\begin{equation}
 0 < \epsilon < \epsilon_0,
\end{equation}
the following statement holds.

For all Riemannian integrable function 
$f(\theta^1, \cdot \cdot \cdot, \theta^N)$ 
which is $2 \pi$-periodic in 
$\theta = (\theta^1, \cdot \cdot \cdot ,\theta^N)$,
the equality
\begin{equation}
 \lim_{T \rightarrow \infty} {1 \over T - 1} \int^T_1 dt f(\theta(t))
 = \overline{f},
\end{equation}
holds where 
\begin{equation}
 \overline{f} = {1 \over (2 \pi)^N} \int^{2 \pi}_0 \cdot \cdot \cdot
 \int^{2 \pi}_0 d^N \theta f(\theta).
\end{equation}

We prove this proposition in the appendix.
As a collorary of the proposition, we obtain

{\bf Corollary}

Under the assumtions in the above proposition, 
as for a definition function $\phi (\theta)$ of the 
Jordan measurable set $A \subset \prod^N_{i=1} [0, 2 \pi]$;
\begin{equation}
 \phi (\theta) = 
   \cases{
      1 & ($\theta \in A$) \cr
      0 & ($\theta \in \prod^N_{i=1} [0, 2 \pi] \setminus A$) \cr
         }
\end{equation}
the equality 
\begin{equation}
 \lim_{T \rightarrow \infty} {1 \over T - 1} \int^T_1 dt 
 \phi(\theta(t)) = {1 \over (2 \pi)^N} \tau (A)
\end{equation}
holds where $\tau (A)$ is a measure of the set $A$.

{\it Proof}

When the set $A$ is Jordan measurable, its definition function 
$\phi (\theta)$ is Riemannian integrable.
Applying the above proposition to $\phi (\theta)$, we obtain 
\begin{eqnarray}
 \lim_{T \rightarrow \infty} {1 \over T - 1} \int^T_1 dt 
 \phi(\theta(t)) &=&
 {1 \over (2 \pi)^N} \int^{2 \pi}_0 \cdot \cdot \cdot 
 \int^{2 \pi}_0 d^N \theta \phi(\theta) 
\nonumber \\
 &=&
 {1 \over (2 \pi)^N} \tau (A).
\end{eqnarray}

{\it Proof End}

From the above corollary, it can be seen that when the masses
of the scalar fields are irrationally independent, 
each track distributes uniformly in the torus $T^N$,
more accurately saying, the period for which an arbitrary track
stays in an arbitrary domain $A$ is proportional to the measure
of $A$. 
Let us adopt the domain which contains 
the singular point $\theta^i = 0$ or $\pi$ 
$(i=1, \cdot \cdot \cdot, N)$ as $A$.  
Then we can evaluate how long the Bardeen parameter becomes 
singular.
But we cannot say when the Bardeen parameter becomes singular.
The occurrence of singularities is essentially probabilistic and 
random.
We can find no rule in a sequence of times when 
$Z$ becomes singular.

\section{Summary and Discussion}

During the reheating, the multiple scalar fields oscillate
arond the local minimum of the potential.
In order to understand the evolutionary behavior of 
the cosmological perturbation on superhorizon scales
during reheating, we investigate that 
in a universe dominated by multiple oscillating scalar fields. 
The dynamical behavior of the cosmological perturbations 
on superhorizon scales is well approximated in terms of that 
of the long wavelength limit, and the latter can be constructed
from the evolution of corresponding exactly homogeneous universe.
These facts provide us with a strong tool to investigate 
the evolution of the cosmological perturbation on superhorizon 
scales including the conservation of the Bardeen parameter.
In order to solve the evolutions of the cosmological 
perturbations in the long wavelength limit, we need know 
the evolution of the exactly homogeneous system.
Since the scalar fields oscillate rapidly, we use the action
angle variables.
When this problem is formulated in the hamiltonian form,
as the result of the expansion of the universe and the interactions
between the scalar fields the perturbed part depending on the angle
variable appears.
We found that it is bounded by the negative power of the time and 
that by choosing the canonical transformation properly,
the perturbed part, in particular the power of the time,
can be decreased so that the transformed hamiltonian can be solved
easily.
By using the long wavelength limit method, and the technique
of the canonical perturbation theory, we prove the fact that
if the ratio of frequncies of the scalar fields are incommensurable,
for the adiabatic growing mode Bardeen parameter stays constant 
and for the other $(2 N - 1)$ modes, the Bardeen parameter decays. 
In addition we show that the Bardeen parameter diverges 
probabilistically.
From the viewpoint of the ergodic theory, we evaluate the period
for which the Bardeen parameter diverges.

In this paper, since the evolution of the cosmological perturbations
are assumed to be generated from the perturbations of the exactly 
homogeneous solutions, we can only take into account the interactions
between the homogenous modes, although in the realistic reheating 
scenario
the interactions between the non-homogeneous modes cannot be neglected and 
may be rather strong.
When we think that the analysis of the latter are too complicated to be 
treated, the investigation of this paper are important, since we can 
investigate analytically, and does not specify the details of 
the interactions.
Although the analysis are confined in the non-resonant case,
the elimination of fast phases by the canonical transformation 
is thought to be still effective also in resonance case.
The analysis in resonant case is in future publications. 

Our final goal is understanding of the evolutionary behaviors of 
the cosmological perturbations during reheating.
We have shown the constancy of the Bardeen parameter in two extreme cases,
that is, in the nearly integrable case corresponding to this paper,
and the chaotic case where the discription 
by the two component fluid model \cite{Hamazaki.T&Kodama1996}
becomes good.
So in order to achieve our purpose completely, general numerical 
calculations
taking into account the interactions between the non-homogeneous modes,
taking its stands between the above two extreme cases are necessary.

\newpage

\section{ Appendix; Mathematical Preliminary }

We give a little knowledge of the complex analysis and 
the fourier analysis,
since it is used in the physical investigations
of this paper.
These propositions are fundamental when we evaluate the upper bound 
of the generating functions of the canonical transformations 
which decrease the size and the power of the time of the perturbed
part of the hamiltonian.

The following proposition is often used when we evaluate 
the derivative with respect to the canonical variables, 
for example the defference between the original variables 
and the transformed variables and the coefficients appearing 
in the perturbations of the canonical equations of motion.

{\bf Proposition}

$f(x)$ is analytic and the inequality
\begin{equation}
 |f(x)| \le C
\end{equation}
holds for some positive $C$ on the domain $D$.
Then for an arbitrary positive $\delta$, in the domain 
$D - \delta = \{x \in D; {\rm ~the~} \delta {\rm ~neighborhood~of~} x 
{\rm ~is~contained~in~} D\}$, the inequalities 
\begin{eqnarray}
&&
 | {\partial f \over \partial x} | \le {C \over \delta},
\\
&&
 | {\partial^2 f \over \partial x^2} | \le {2 C \over \delta^2},
\end{eqnarray}
hold.

{\it Proof}

Use the Cauchy equation
\begin{equation}
f(x) = {1 \over 2 \pi i} \oint dy {f(y) \over y-x}.
\end{equation}
{\it Proof End}

The following proposition guarantee the fact that 
the derivative with respect to the time decreases the power
of the time.

{\bf Proposition}

$t^m S(t)$ is analytic function of $s = {1 / t}$ 
on the domain $s \in C(\rho)$ where
\begin{equation}
 C(\rho)=\{ s; |s|\le 1+\rho, |{\rm Im}s| \le \rho \}
\end{equation}
and the inequality
\begin{equation}
 |t^m S| \le C
\end{equation}
holds for some positive $C$.
Then for an arbitrary positive $\delta$, on the domain $C(\rho - \delta)$ 
$t^{m+1} {\partial S / \partial t}$ is analytic and the inequality 
\begin{equation}
 |t^{m+1} {\partial S \over \partial t}| \le C (m + {1 + \rho \over \delta})
\end{equation} 
holds.

{\it Proof}

By differentiating the equation
\begin{equation}
{1 \over s^m} S({1 \over s}) = \oint dz {1 \over z-s} 
{1 \over z^m} S({1 \over z}),
\end{equation}
we obtain 
\begin{equation}
 {1 \over s^{m+1}} S'({1 \over s}) = 
 - m {1 \over s^m} S({1 \over s})- {s \over 2 \pi i}
 \oint dz {1 \over (z-s)^2} 
 {1 \over z^m} S({1 \over z}),
\end{equation}
from which the proposition follows.

{\it Proof End}

{\bf Proposition}

$F(\theta) = \sum_k F_k \exp{i k \cdot \theta}$ 
is periodic in $\theta = (\theta_1, \cdot \cdot \cdot, \theta_N)$
and analytic and the inequality
\begin{equation}
 |F(\theta)| \le C
\end{equation}
holds for some positive $C$ on the domain
\begin{equation}
 |{\rm Im}\theta| \le \rho.
\end{equation}
Then the fourier components $F_k$ satisfy
\begin{equation}
 | F_k | \le C \exp (- |k| \rho)
\end{equation}
where $|k|$ is defined by
\begin{equation}
 |k| = |k_1| + \cdot \cdot \cdot + |k_N|.
\end{equation}

{\it Proof}

The fourier coefficients are given by
\begin{equation}
 F_k = {1 \over (2 \pi)^N} \int^{2 \pi}_0 \cdot \cdot \cdot
 \int^{2 \pi}_0 d\theta_1 \cdot \cdot \cdot d\theta_N 
 \exp(- i k \cdot \theta) F(\theta).
\end{equation}
By changing the contour into $\theta_i = x_i + i \rho_i$,
($0 \le x_i \le 2 \pi$, $\rho_i = - {\rm sgn}(k_i) \rho$)
taking into account that contributions from the both sides 
cancel,
$F_k$ is evaluated as 
\begin{eqnarray}
 |F_k| &\le& \prod_i \exp(- |k_i| \rho)C
\nonumber \\
 &=& C \exp(- |k| \rho).
\end{eqnarray}

{\it Proof End}

The following proposition guarantees that sufficiently small
generating function defines the diffeomorphic 
canonical transformations.
We omit the proof.

{\bf Proposition}

Suppose that the mapping $T; x \rightarrow y$ is defined 
on the domain $D$.
When $T$ is differentiable and 
${\rm det} {\partial y / \partial x} \ne 0$ and 
$|x - y| \le \epsilon$ on the domain $D$,
$T$ is diffeomorphic on the domain $D - 4 \epsilon$.

\section{ Appendix; Ergodic Property
 of Conditionally Periodic Motion }

In chapter 6, we pointed out that the Bardeen parameter becomes
singular at a certain point in the torus $T^N$.
In the viewpoint of the ergodic theory, we cannot say when the 
Bardeen parameter becomes singular, but can say how long 
the Bardeen parameter singular.
In this appendix, we discuss the ergodic property as for the 
universe dominated by multiple oscillating scalar fields
whose masses are incommensurable, i.e. irrationally independent.
In our system, it can be shown that an arbitrary track is dense 
everywhere in the torus $T^N$.
It is referred to as a conditionary periodic motion.

We consider the system proposed 
in Eq.(\ref {startinghamiltonian}).
We assume that $\mu$ is a linearly independent vector on 
${\bf Z}^N$;
\begin{equation}
 \mu \cdot k = 0 \Rightarrow k = 0.
\end{equation}
As a first step, we prove the lemma below.

{\bf Lemma}

\begin{equation}
 \lim_{T \rightarrow \infty} {1 \over T - 1} \int^T_1 dt 
 e^{i k \cdot \theta(t)} = 0,
\end{equation}

{\it Proof}

We consider the quantity as
\begin{equation}
 \int_1 dt F(J, \theta, t).
\end{equation}
If $G(J, \theta, t)$ satisfying 
\begin{equation}
 {1 \over \epsilon} {1 \over (\mu \cdot J)^{1 / 2}}
 \sum_i \mu_i {\partial G \over \partial \theta^i} = F,
\end{equation}
exists, the equality
\begin{equation}
 \int_1 dt F(J, \theta, t) =
 [ G(J, \theta, t) ]_{t=1} +
 \int_1 dt [
 - \sum_i 
 {\partial {\tilde H} \over \partial J^i} 
 {\partial G \over \partial \theta^i} 
 + \sum_i    
 {\partial {\tilde H} \over \partial \theta^i} 
 {\partial G \over \partial J^i} 
  - {\partial G \over \partial t}
           ]
\end{equation}
holds.
When $F = e^{i k \cdot \theta}$, 
$G = {\epsilon / i (\mu \cdot k)} \cdot
(\mu \cdot J)^{1 / 2} e^{i k \cdot \theta}$.
Then there exists positive $M$, $N$ such that 
\begin{equation}
 |{1 \over T-1} \int^T_1 dt e^{i k \cdot \theta}| \le 
 {\epsilon \over (\mu \cdot k)} {1 \over T-1} [2 M + N \ln T]
 \rightarrow 0 \quad
 (T \rightarrow \infty).
\end{equation}
We complete the proof.

{\it Proof End}

{\bf Lemma}

For a triangular polynomial 
\begin{equation}
 P_M (\theta) = \sum_{|k| \le M} f_k e^{i k \cdot \theta},
\end{equation}
the equality
\begin{equation}
 \lim_{T \rightarrow \infty} {1 \over T - 1} \int^T_1 dt 
 P_M (\theta) = \overline{ P_M }
\label {polynomialergodic}
\end{equation}
holds.

{\it Proof}

The both sides of Eq.(\ref {polynomialergodic}) is linear.
Therefore from the above lemma, Eq.(\ref {polynomialergodic})
holds.

{\it Proof End}

{\bf Lemma; Weierstrass' theorem}

Suppose that $f({\bf x})$ is continuous on 
$\prod^N_{i=1} [a_i, b_i]$.
For an arbitrary positive $\epsilon$, there exists a polynomial
$P({\bf x})$ such that 
\begin{equation}
 |f({\bf x})-P({\bf x})| < \epsilon \quad \quad
 {\bf x} \in \prod^N_{i=1} [a_i, b_i].
\end{equation}

{\it Proof}

We consider the integral;
\begin{eqnarray}
 &&
 J_n = \int^1_0 (1 - v^2)^n dv,
\\
 &&
 J_n^* = \int^1_\delta (1 - v^2)^n dv \quad
 (0 < \delta < 1).
\end{eqnarray}
Since 
\begin{eqnarray}
 &&
 J_n > \int^1_0 (1-v)^n dv = {1 \over n+1},
\\
 &&
 J_n^* < (1 - \delta^2)^n (1-\delta) < (1 - \delta^2)^n,
\\
 &&
 {J_n^* \over J_n} < (n+1) (1 - \delta^2)^n,
\end{eqnarray}
we obtain 
\begin{equation}
 \lim_{n \rightarrow \infty} {J_n^* \over J_n} = 0. 
\label {limitJn}
\end{equation}
Without loss of generality, we can put $0 < a_i < b_i < 1$.
We insert $\alpha_i$, $\beta_i$ such that 
$0 < \alpha_i < a_i < b_i < \beta_i < 1$.
As a polynomial, we adopt 
\begin{equation}
 P_n ({\bf x}) = {1 \over 2^N (J_n)^N} \int_C f({\bf u})
 [1 - (u_1 - x_1)^2]^n \cdot \cdot \cdot [1 - (u_N - x_N)^2]^n
 d u_1 \cdot \cdot \cdot d u_N,
\end{equation}
where
\begin{equation}
 C = \prod^N_{i=1} [\alpha_i, \beta_i].
\end{equation}
$P_n ({\bf x})$ is a polynomial of $2 n$ degree in terms of $x_i$,
($i=1, \cdot \cdot \cdot, N$).
We evaluate how accurately the polynomial $P_n ({\bf x})$ approximates
the continuous function $f ({\bf x})$.
We introduce sets $A$, $B$ defined by
\begin{eqnarray}
&&
 A = \prod^N_{i=1} [x_i-\delta, x_i+\delta] \quad \quad 
 (0 < \delta < 1),
\\
&&
 B = \prod^N_{i=1} [a_i, b_i].
\end{eqnarray}
First we evaluate
\begin{eqnarray}
 I \equiv && \int_A f({\bf u})
 [1 - (u_1 - x_1)^2]^n \cdot \cdot \cdot [1 - (u_N - x_N)^2]^n
 d u_1 \cdot \cdot \cdot d u_N 
\nonumber \\
 && - 2^N (J_n)^N f({\bf x}).
\end{eqnarray}
In terms of ${\bf v} = {\bf u}- {\bf x}$, $A$ is transferred into
$A'$ defined by
\begin{equation}
 A' = \prod^N_{i=1} [-\delta, \delta].
\end{equation}
Therefore we obtain
\begin{eqnarray}
 I = && \int_{A'} f({\bf x})
 [1 - (u_1 - x_1)^2]^n \cdot \cdot \cdot [1 - (u_N - x_N)^2]^n
 d u_1 \cdot \cdot \cdot d u_N 
\nonumber \\
 && + \int_{A'} ( f({\bf v}+{\bf x})-f({\bf x}) )
 [1 - (u_1 - x_1)^2]^n \cdot \cdot \cdot [1 - (u_N - x_N)^2]^n
 d u_1 \cdot \cdot \cdot d u_N 
\nonumber \\
 && - 2^N (J_n)^N f({\bf x}).
\end{eqnarray}
Since $f({\bf x})$ is uniformly continuous on ${\bf x} \in B$,
there exists positive $\delta$ which depends only on $\epsilon$
($0 < \delta < 1$) such that 
\begin{equation}
 |{\bf v}| \le \sqrt{m} \delta
\end{equation}
and
\begin{equation}
 | f({\bf v}+{\bf x})-f({\bf x}) | \le \epsilon \quad \quad
 ({\bf x} \in B).
\end{equation}
Therefore we obtain
\begin{equation}
 |{\rm I}| \le 2^N \{    
 (J_n)^N - (J_n - J_n^*)^N
             \} M
 + \epsilon 2^N (J_n - J_n^*)^N,
\label {evaluateI}
\end{equation}
where
\begin{equation}
 M = \sup_{{\bf x} \in C} |f({\bf x})|.
\end{equation}
Next we evaluate 
\begin{equation}
 {\rm II} = \int_{C \setminus A} f({\bf u})
 [1 - (u_1 - x_1)^2]^n \cdot \cdot \cdot [1 - (u_N - x_N)^2]^n
 d u_1 \cdot \cdot \cdot d u_N. 
\end{equation}
Since in terms of ${\bf v} = {\bf u} - {\bf x}$, $C$, $A$ are 
transferred into $C'$, $A'$ defined by
\begin{eqnarray}
 &&
 C' = \prod^N_{i=1} [\alpha_i - x_i, \beta_i - x_i],
\nonumber \\
 &&
 A' = \prod^N_{i=1} [-\delta, \delta],
\end{eqnarray}
and 
\begin{equation}
 C'\subset D=\prod^N_{i=1} [-1, 1],
\end{equation}
we obtain 
\begin{equation}
 |{\rm II}| \le M 2^N \{ 
 (J_n)^N - (J_n - J_n^*)^N
                \}.
\label {evaluateII}
\end{equation}
From Eqs.(\ref {evaluateI}), (\ref {evaluateII}), we obtain
\begin{equation}
 |P_n ({\bf x})- f ({\bf x})| \le
 |{\rm I}| + |{\rm II}| \le
 2 M \{ 1 - (1-{J_n^* \over J_n})^N \} 
 + \epsilon (1-{J_n^* \over J_n})^N.
\end{equation}
From Eq.(\ref {limitJn}), the right hand side can be smaller 
than $2 \epsilon$ when we adopt sufficiently large $n$.
We complete the proof.

{\it Proof End}

{\bf Lemma}
 
Suppose that $f (\theta^1, \cdot \cdot \cdot, \theta^N)$ is 
continuous and $2 \pi$-periodic.
For an arbitrary positive $\epsilon$, there exists a triangular
polynomial $P(\theta^1, \cdot \cdot \cdot, \theta^N)$;
\begin{equation}
 P(\theta^1, \cdot \cdot \cdot, \theta^N) = {\alpha_0 \over 2}
 + \sum_{|k| \le M} \alpha_k e^{i k \cdot \theta},
\end{equation}
such that
\begin{equation}
 |f(\theta) - P(\theta)| < \epsilon.
\end{equation}

{\it Proof}

We define $\phi (\xi, \eta)$ by
\begin{equation}
 \phi (\xi, \eta) = \rho_1 \cdot \cdot \cdot \rho_N 
      f (\theta^1, \cdot \cdot \cdot, \theta^N),
\end{equation}
where 
\begin{eqnarray}
 &&
 \xi_i = \rho_i \cos \theta^i,
\\
 &&
 \eta_i = \rho_i \sin \theta^i.
\end{eqnarray}
When $\rho^2_i = \xi^2_i + \eta^2_i = 1$, 
($i=1, \cdot \cdot \cdot, N$), $\phi (\xi, \eta)$ coincide
with $f (\theta)$.
Since $\phi (\xi, \eta)$ is continuous for $\xi_i$, $\eta_i$,
from Weierstrass' theorem, there exists a polynomial for 
$\xi$, $\eta$ $P_n (\xi, \eta)$ which approximates uniformly
$\phi (\xi, \eta)$ in a rectangle containing 
$\xi^2_i + \eta^2_i = 1$ ($i=1, \cdot \cdot \cdot, N$).
This completes the proof.

{\it Proof End}

{\bf Proposition}

For an arbitrary Riemannian integrable function $f (\theta)$,
the equality 
\begin{equation}
 \lim_{T \rightarrow \infty} {1 \over T-1} \int^T_1 dt
 f (\theta(t)) = \overline{f}
\end{equation}
holds.

{\it Proof}

For an arbitrary positive $\epsilon$, there exists triangular
polynomials $P_1$, $P_2$ such that
\begin{eqnarray}
&&
 P_1 < f < P_2,
\label {approximationbypolynomial}
\\
&&
 \overline{P_2} - \overline{P_1} < \epsilon. 
\end{eqnarray}
From Eq. (\ref {approximationbypolynomial}), we obtain
\begin{equation}
 {1 \over T-1} \int^T_1 P_1 (\theta(t)) dt <
 {1 \over T-1} \int^T_1 f (\theta(t)) dt <
 {1 \over T-1} \int^T_1 P_2 (\theta(t)) dt.
\end{equation}
From the above inequality,
by using the ergodic equality of the triangular polynomial,
we obtain
\begin{eqnarray}  
 \overline{P_1} &\le& \lim_{T_0 \rightarrow \infty}
 \inf_{T \ge T_0}{1 \over T-1} \int^T_1 f (\theta(t)) dt
\nonumber \\
 &\le& \lim_{T_0 \rightarrow \infty}
 \sup_{T \ge T_0}{1 \over T-1} \int^T_1 f (\theta(t)) dt
 \le \overline{P_2}
\end{eqnarray} 
which yields
\begin{equation}
 \lim_{T_0 \rightarrow \infty}
 \sup_{T \ge T_0}{1 \over T-1} \int^T_1 f (\theta(t)) dt -
 \lim_{T_0 \rightarrow \infty}
 \inf_{T \ge T_0}{1 \over T-1} \int^T_1 f (\theta(t)) dt
 < \epsilon.
\end{equation}

Since $\epsilon$ is arbitrary, and 
$\overline{P_1} < \overline{f} < \overline{P_2}$,
we obtain
\begin{equation}  
 \lim_{T \rightarrow \infty} {1 \over T-1}
 \int^T_1 f(\theta(t)) dt = \overline{f}.
\end{equation}  
We complete the proof.

{\it Proof End}

\section*{Acknowledgments}

The author would like to thank Professor H.Kodama for valuable discussions
and comments, without which this paper would not have emerged. 
He would like to thank Professor H.Ito for giving him information 
about the hamiltonian dynamics. 
He would like to thank Professor A.Hosoya for continuous encouragements.


\addtolength{\baselineskip}{-3mm}

\end{document}